\documentclass[aps,prl,superscriptaddress,showpacs]{revtex4}

\usepackage[dvips]{graphicx}
\usepackage[dvips]{color}
\usepackage{epsfig}
\usepackage{amssymb,amsmath}
\usepackage{lscape}

\newcommand{\bdk}{$B^{\pm}\to DK^{\pm}$}
\newcommand{\bdtk}{$B^{\pm}\to \tilde{D}K^{\pm}$}

\newcommand{\bdsk}{$B^{\pm}\to D^{*}K^{\pm}$}
\newcommand{\bdstk}{$B^{\pm}\to \tilde{D}^{*}K^{\pm}$}

\newcommand{\bddsk}{$B^{\pm}\to D^{(*)}K^{\pm}$}
\newcommand{\bddstk}{$B^{\pm}\to \tilde{D}^{(*)}K^{\pm}$}

\newcommand{\bdpi}{$B^{\pm}\to D\pi^{\pm}$}
\newcommand{\bdtpi}{$B^{\pm}\to \tilde{D}\pi^{\pm}$}

\newcommand{\bdspi}{$B^{\pm}\to D^{*}\pi^{\pm}$}
\newcommand{\bdstpi}{$B^{\pm}\to \tilde{D}^{*}\pi^{\pm}$}

\newcommand{\bddspi}{$B^{\pm}\to D^{(*)}\pi^{\pm}$}
\newcommand{\bddstpi}{$B^{\pm}\to \tilde{D}^{(*)}\pi^{\pm}$}

\newcommand{\dsdpi}{$D^{*\pm}\to D\pi^{\pm}$}
\newcommand{\dsdpis}{$D^{*\pm}\to D\pi_s^{\pm}$}

\newcommand{\dkpp}{$\bar{D^0}\to K_S\pi^+\pi^-$}
\newcommand{\dtkpp}{$\tilde{D}\to K_S\pi^+\pi^-$}

\begin{document}

\title{Measurement of \boldmath{$\phi_3$} with Dalitz Plot
Analysis of \boldmath{\bddsk} Decay}

\affiliation{Budker Institute of Nuclear Physics, Novosibirsk}
\affiliation{Chiba University, Chiba}
\affiliation{Chonnam National University, Kwangju}
\affiliation{University of Cincinnati, Cincinnati, Ohio 45221}
\affiliation{Gyeongsang National University, Chinju}
\affiliation{University of Hawaii, Honolulu, Hawaii 96822}
\affiliation{High Energy Accelerator Research Organization (KEK), Tsukuba}
\affiliation{Hiroshima Institute of Technology, Hiroshima}
\affiliation{Institute of High Energy Physics, Chinese Academy of Sciences, Beijing}
\affiliation{Institute of High Energy Physics, Vienna}
\affiliation{Institute for Theoretical and Experimental Physics, Moscow}
\affiliation{J. Stefan Institute, Ljubljana}
\affiliation{Kanagawa University, Yokohama}
\affiliation{Korea University, Seoul}
\affiliation{Kyungpook National University, Taegu}
\affiliation{Swiss Federal Institute of Technology of Lausanne, EPFL, Lausanne}
\affiliation{University of Ljubljana, Ljubljana}
\affiliation{University of Maribor, Maribor}
\affiliation{University of Melbourne, Victoria}
\affiliation{Nagoya University, Nagoya}
\affiliation{Nara Women's University, Nara}
\affiliation{National Kaohsiung Normal University, Kaohsiung}
\affiliation{National United University, Miao Li}
\affiliation{Department of Physics, National Taiwan University, Taipei}
\affiliation{H. Niewodniczanski Institute of Nuclear Physics, Krakow}
\affiliation{Nihon Dental College, Niigata}
\affiliation{Niigata University, Niigata}
\affiliation{Osaka City University, Osaka}
\affiliation{Osaka University, Osaka}
\affiliation{Panjab University, Chandigarh}
\affiliation{Peking University, Beijing}
\affiliation{Princeton University, Princeton, New Jersey 08545}
\affiliation{Saga University, Saga}
\affiliation{University of Science and Technology of China, Hefei}
\affiliation{Seoul National University, Seoul}
\affiliation{Sungkyunkwan University, Suwon}
\affiliation{University of Sydney, Sydney NSW}
\affiliation{Tata Institute of Fundamental Research, Bombay}
\affiliation{Toho University, Funabashi}
\affiliation{Tohoku Gakuin University, Tagajo}
\affiliation{Tohoku University, Sendai}
\affiliation{Department of Physics, University of Tokyo, Tokyo}
\affiliation{Tokyo Institute of Technology, Tokyo}
\affiliation{Tokyo Metropolitan University, Tokyo}
\affiliation{Tokyo University of Agriculture and Technology, Tokyo}
\affiliation{University of Tsukuba, Tsukuba}
\affiliation{Virginia Polytechnic Institute and State University, Blacksburg, Virginia 24061}
\affiliation{Yonsei University, Seoul}
  \author{A.~Poluektov}\affiliation{Budker Institute of Nuclear Physics, Novosibirsk} 
  \author{K.~Abe}\affiliation{High Energy Accelerator Research Organization (KEK), Tsukuba} 
  \author{T.~Abe}\affiliation{High Energy Accelerator Research Organization (KEK), Tsukuba} 
  \author{H.~Aihara}\affiliation{Department of Physics, University of Tokyo, Tokyo} 
  \author{Y.~Asano}\affiliation{University of Tsukuba, Tsukuba} 
  \author{T.~Aushev}\affiliation{Institute for Theoretical and Experimental Physics, Moscow} 
  \author{T.~Aziz}\affiliation{Tata Institute of Fundamental Research, Bombay} 
  \author{S.~Bahinipati}\affiliation{University of Cincinnati, Cincinnati, Ohio 45221} 
  \author{A.~M.~Bakich}\affiliation{University of Sydney, Sydney NSW} 
  \author{I.~Bedny}\affiliation{Budker Institute of Nuclear Physics, Novosibirsk} 
% ...
\collaboration{The Belle Collaboration}

\begin{abstract} 

We present a measurement of the unitarity triangle angle $\phi_3$
using a Dalitz plot analysis of the three-body decay of the neutral 
$D$ meson from the \bddsk\ process. 
The method employs the interference 
between $D^0$ and $\bar{D^0}$ to extract both the weak and strong phases. 
We apply this method to a 140 fb$^{-1}$ data sample
collected by the Belle experiment. 
The analysis uses the modes \bdk\ and \bdsk, $D^{*}\to D\pi^0$, where
the neutral $D$ meson decays into $K_S\pi^+\pi^-$. 
We obtain 146 signal candidates for \bdk\ and 
39 candidates for \bdsk.
From a combined maximum likelihood fit to the \bdk\ and \bdsk\ modes, 
we obtain $\phi_3=77^{\circ}\;^{+17^{\circ}}_{-19^{\circ}}
\mbox{(stat)}\pm 13^{\circ}
\mbox{(syst)}\pm 11^{\circ}(\mbox{model})$. The
corresponding two standard deviation 
interval is $26^{\circ}<\phi_3<126^{\circ}$. 
\end{abstract}
\maketitle

\section{Introduction}

\begin{figure}
  \begin{center}
  \epsfig{figure=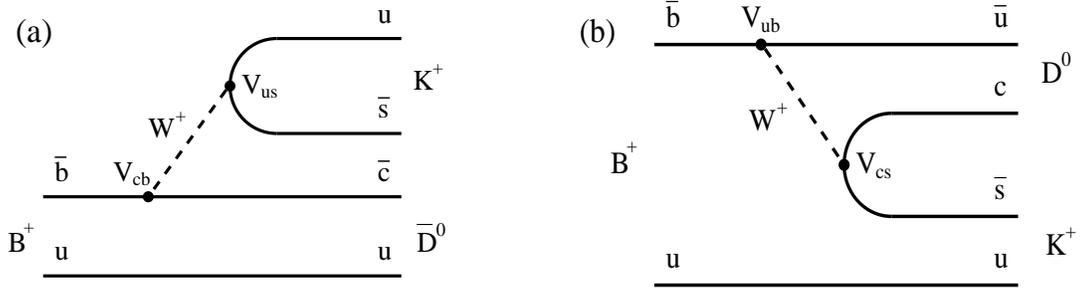,width=0.8\textwidth}
  \caption{Feynman diagrams of (a) dominant $B^+\to \bar{D^0}K^+$ and 
           (b) suppressed $B^+\to D^0K^+$ decays}
  \label{diags}
  \end{center}
\end{figure}

Determinations of the Cabbibo-Kobayashi-Maskawa
(CKM) \cite{ckm} matrix elements provide important checks on
the consistency of the Standard Model and ways to search
for new physics. The possibility of observing direct CP violation
in $B\to D K$ decays
was first discussed by I. Bigi and A. Sanda \cite{bigi}.
Since then, various methods using CP violation in $B\to D K$ decays have been
proposed \cite{glw,dunietz,eilam,ads} to measure the unitarity triangle
angle $\phi_3$. These methods are based on two key observations:
neutral $D^{0}$ and $\bar{D^0}$
mesons can decay to a common final state, and the decay
$B^+\to D^{(*)} K^+$ can produce neutral $D$ mesons of both flavors
via $\bar{b}\to \bar{c}u\bar{s}$ (Fig.~\ref{diags}a)
and $\bar{b}\to \bar{u}c\bar{s}$ (Fig.~\ref{diags}b) transitions,
with a relative phase $\theta_+$ between the two interfering
amplitudes that is the sum, $\delta + \phi_3$, of strong and weak interaction
phases.  For the charge conjugate mode, the relative phase is
$\theta_-=\delta-\phi_3$, so both phases can be extracted
from measurements of such charge conjugate $B$ decay modes.
However, the use of branching fractions alone requires additional
information to obtain $\phi_3$.
This is provided either by determining the branching fractions of
decays to flavour eigenstates (GLW method \cite{glw})
or by using different neutral $D$ final states (ADS method \cite{ads}).

A Dalitz plot analysis of a three-body final state of the $D$ meson
allows one to obtain all the information required for determination
of $\phi_3$ in a single decay mode. The use of a Dalitz plot analysis
for the extraction of $\phi_3$ was first discussed
by D. Atwood, I. Dunietz and A. Soni in application to the ADS
method \cite{ads}. This technique uses the interference of
Cabibbo-favored $D^0\to K^+\pi^-\pi^0$ and doubly Cabibbo-suppressed
$\bar{D^0}\to K^+\pi^-\pi^0$ decays.
However, the small rate for the doubly Cabibbo-suppressed decay
limits the experimental applicability of this technique.
Recently, three body final states common to $D^0$ and
$\bar{D^0}$, such as $K_S\pi^+\pi^-$ \cite{giri}, were suggested as
being more promising,
since both interfering amplitudes are Cabibbo-favored in this
case. This technique appears to have a higher statistical
precision than methods based on branching fraction measurements.
The statistical accuracy of the $\phi_3$ extraction can be improved
by adding the excited states of $D$ and $K$ to the analysis \cite{atwood}.

In the Wolfenstein parameterization of the CKM matrix elements, 
the amplitudes of the two diagrams
that contribute to the decay $B^+\to D K^+$
are given by $M_1\sim V_{cb}^*V_{us}\sim A\lambda^3$ 
(for the $\bar{D^0} K^+$ final state) and
$M_2\sim V_{ub}^*V_{cs}\sim A\lambda^3(\rho+i\eta)$
(for $D^0 K^+$).
The annihilation diagram also contributes to $M_2$, but, 
since the weak coefficients are the same, this effectively leads
to a redefinition of the strong phase. The two amplitudes 
$M_1$ and $M_2$ interfere as the $D^0$ and $\bar{D^0}$ mesons decay
into the same final state $K_S \pi^+ \pi^-$; 
we denote the admixed state as $\tilde{D}$. Assuming no $CP$ 
asymmetry in neutral $D$ decays, the amplitude of the $B^+$ decay can be written as
\begin{equation}
\label{intdist}
M_+=f(m^2_+, m^2_-)+re^{i\phi_3+i\delta}f(m^2_-, m^2_+), 
\end{equation}
where $m_{+}^2$ and $m_{-}^2$ are the squared 
invariant masses of the $K_S \pi^+$ and
$K_S \pi^-$ combinations, respectively, and $f(m_+, m_-)$ 
is the complex amplitude for the decay
\dkpp. The absolute value of the ratio between 
the two interfering amplitudes, $r$, is given by the ratio
$|V_{ub}^*V_{cs}|/|V_{cb}^*V_{us}|\sim 0.38$ and
the color suppression factor. The latter can be roughly estimated 
from the ratio of the color suppressed $\bar{B^0}\to D^0 \bar{K^0}$
\cite{bdksupp_belle} and color allowed $B^-\to D^0 K^-$ decays
\cite{bdk_belle}:
$\sqrt{\mathcal{B}(\bar{B^0}\to D^0 \bar{K^0})/\mathcal{B}(B^-\to D^0 K^-)}=0.35\pm 0.05$.
The amplitude ratio is therefore expected to be 
the product of these two factors, {\it i.~e.} $r\sim 0.13$. 
However, other estimations of $r$ exist, predicting the values 
as large as 0.2 \cite{gronau}. 

The corresponding amplitude for the charge conjugate $B^-$ decay is 
\begin{equation}
\label{intdist_m}
M_-=f(m^2_-, m^2_+)+re^{-i\phi_3+i\delta}f(m^2_+, m^2_-). 
\end{equation}
Once the functional form of $f$ is fixed by a model for \dkpp, 
the $\tilde{D}$ Dalitz distributions for $B^+$ and $B^-$ decays can be 
fitted simultaneously using the above expressions for $M_{+}$ and $M_{-}$, 
with $r$, $\phi_3$, and $\delta$ as free parameters.
There are certain advantages of this technique:
it is directly sensitive to the value of $\phi_3$ and 
does not require additional assumptions on the values of $r$ and $\delta$. 
Moreover, the value of $r$ obtained in the fit can be useful for
other $\phi_3$ measurements. 

Reference \cite{giri} suggests a model-independent way for determining 
$\phi_3$ via a binned Dalitz plot analysis. However, the 
application of this procedure would result in 
large statistical errors with our present data sample. 
Instead, we use a model-dependent approach based on an unbinned maximum likelihood
fit to the \dtkpp\ Dalitz plot distributions corresponding to 
$B^+$ and $B^-$ data samples, thus making optimal use of our small number of 
signal events. 
The model of \dkpp\ decay in our approach is determined
from a large sample of flavor-tagged \dkpp\ decays 
produced in continuum $e^+e^-$ annihilation.
The drawback of this approach is that only the absolute value of the 
$D^0$ decay amplitude $f$ can be determined directly; the complex 
form of $f$ has to be based on model assumptions; 
these lead to potential model-dependent uncertainties
in the determination of $\phi_3$. 
Note, however, that the model uncertainties can be controlled in the 
future using data from $\tau$-charm factories. $CP$ tagged neutral $D$ mesons
can be produced in the decay of the $\psi(3770)$ resonance, and these 
can be used to obtain information about the complex phase of the 
amplitude $f$, which is precisely the information required for a 
model-independent measurement of $\phi_3$.

The method used here has two possible two-fold ambiguities in the determination 
of the pair of parameters $(\phi_3, \delta)$. The first one is a shift 
$(\phi_3, \delta)\to (\phi_3+\pi, \delta+\pi)$. The measured phases 
$\theta_+=\delta+\phi_3$ and $\theta_-=\delta-\phi_3$ 
do not change under this transformation. Another ambiguity
is the inversion of sign $(\phi_3, \delta)\to (-\phi_3, -\delta)$
with the simultaneous complex conjugation of the $\bar{D^0}$ decay amplitude $f$. 
This transformation does not change the observables, which are 
the squared absolute values
of the amplitudes. However, if the $\bar{D^0}$ decay amplitude is approximated
by a set of two-body amplitudes, the Breit-Wigner dependence fixes the 
sign of the imaginary part of the $\bar{D^0}$ decay amplitude 
(the complex conjugate Breit-Wigner amplitude does not satisfy the 
causality requirement), and the second ambiguity is thus resolved. 

In a preliminary version of this analysis \cite{belle_dalitz}, 
we used only the \bdtk\ 
mode to constrain $\phi_3$. The current measurement is based on two
modes, \bdtk\ and \bdstk\ ($D^*\to D\pi^0$). 
The statistical error evaluation is also 
improved compared to the previous measurement. 

\section{Event selection}

We use a 140 fb$^{-1}$ data sample collected by 
the Belle detector. The decays \bdk\ and 
\bdsk, $D^{*}\to D\pi^0$ are selected for the 
determination of $\phi_3$; the decays \bdpi, 
\bdspi\ with $D^{*}\to D\pi^0$ 
and $\bar{B^0}(B^0)\to D^{*\pm}\pi^{\mp}$ with $D^{*\pm}\to D\pi^{\pm}$ 
serve as control samples. 
We require the neutral $D$ meson to decay to the 
$K_S\pi^+\pi^-$ final state in all cases.
We also select decays of \dsdpi\ produced via the 
$e^+e^-\to c\bar{c}$ continuum process as a high-statistics 
sample to determine the \dkpp\ decay amplitude. 

The Belle detector is described in detail elsewhere \cite{belle}. 
It is a large-solid-angle magnetic spectrometer consisting of a three-layer
silicon vertex detector (SVD), a 50-layer central drift chamber (CDC) for
charged particle tracking and specific ionization measurement ($dE/dx$), 
an array of aerogel threshold \v{C}erenkov counters (ACC), time-of-flight
scintillation counters (TOF), and an array of 8736 CsI(Tl) crystals for 
electromagnetic calorimetry (ECL) located inside a superconducting solenoid coil
that provides a 1.5 T magnetic field. An iron flux return located outside 
the coil is instrumented to detect $K_L$ mesons and identify muons (KLM).

Separation of kaons and pions is accomplished by combining the responses of 
the ACC and the TOF with the $dE/dx$ measurement from the CDC to form a likelihood 
$\mathcal{L}(h)$ where $h$ is a pion or a kaon. Charged particles are 
identified as pions or kaons using the likelihood ratio
$\mathcal{R}_{\rm PID}(h)  =\mathcal{L}(h)/(\mathcal{L}(K)+\mathcal{L}(\pi))$.

Charged tracks are required to satisfy criteria based on the 
quality of the track fit and the distance from the interaction point in both
longitudinal and transverse planes with respect to the beam axis. 
To reduce the low momentum combinatorial 
background we require each track to have a transverse momentum greater than 
100 MeV/$c$. For charged kaon identification, we require the track to
have $\mathcal{R}_{\rm PID}(K)>0.7$. 

Photon candidates are required to have ECL energy greater than 30 MeV. 
Neutral pion candidates are formed from pairs of photons with invariant 
masses in the range 120 to 150 MeV/$c^2$, or less than two standard 
deviations from the $\pi^0$ mass.

Neutral kaons are reconstructed from pairs of oppositely charged tracks
without any pion PID requirement.
We require the reconstructed vertex distance from the interaction point 
in the plane transverse to the beam axis to be more than 1 mm 
and the invariant mass $M_{\pi\pi}$
to satisfy $|M_{\pi\pi}-M_{K_S}|<10$ MeV/$c^2$, or less than four standard 
deviations from the nominal $K_S$ mass. 


\subsection{Selection of \boldmath{\dsdpi}}

To determine the $\bar{D^0}$ decay model we use $D^{*\pm}$ mesons
produced via the $e^+ e^-\to c\bar{c}$ continuum process. 
The flavor of the neutral $D$ meson is tagged by the charge of the slow pion 
(which we denote as $\pi_s$) in the decay \dsdpis.

To select neutral $D$ candidates we require the invariant mass of the 
$K_S\pi^+\pi^-$ system to be within 9 MeV/$c^2$ of the $D^0$ mass, $M_{D^0}$.
To select events originating from a $D^{*\pm}$ decay 
we make a requirement on the difference 
$\Delta M=M_{K_S\pi^+\pi^-\pi_s}-M_{K_S\pi^+\pi^-}$ of the invariant masses of
the $D^{*\pm}$ and the neutral $D$ candidates: 
$144.6\mbox{ MeV}/c^2<\Delta M<146.4\mbox{ MeV}/c^2$.
To suppress combinatorial background from $B\bar{B}$ events, 
we require the $D^{*\pm}$ to have momentum 
in the center-of-mass (CM) frame greater than 2.7 GeV/$c$.

\begin{figure}
  \epsfig{figure=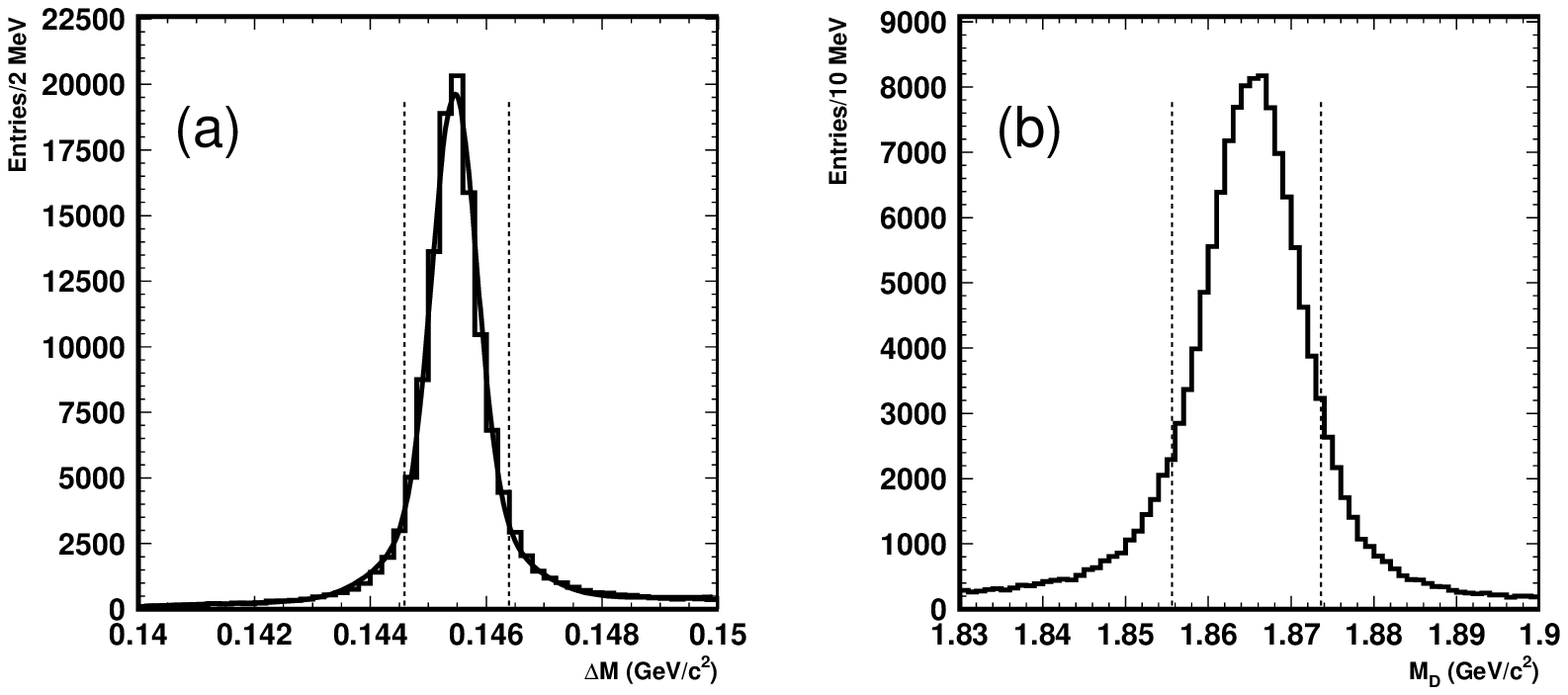,width=\textwidth}
  \caption{(a) $\Delta M$ and (b) $M_{D}$ distributions for the \dsdpis
  candidates. Dashed lines show the signal region. The histogram shows the data;
  the smooth curve in (a) is the fit result. }
  \label{d0_sel}
\end{figure}

The distributions of $\Delta M$ and $M_{K_S\pi^+\pi^-}$ for these 
events are shown in 
Fig.~\ref{d0_sel}. The signal region bounds are indicated with dashed lines. 
The resolutions of the selection variables 
are $\sigma(\Delta M)=0.38$ MeV$/c^2$ and 
$\sigma(M_{K_S\pi^+\pi^-})=5.4$ MeV$/c^2$. The number of events 
that pass all selection criteria is 104204. To obtain the number of background 
events in our sample we fit the $\Delta M$ distribution. The background is 
parameterized with the function 
$b(\Delta M)\sim (1/\Delta M)\sqrt{(\Delta M/m_{\pi})^2-1}$;
the function describing the signal is a combination of two Gaussian peaks with the same 
mean value. The fit yields $100870\pm 840$ signal events and 
$3210\pm 50$ background events
corresponding to a background fraction of 3.1\%.


\subsection{Selection of \boldmath{\bdk}}

The selection of $B$ candidates is based on the CM energy difference
$\Delta E = \sum E_i - E_{\rm beam}$ and the beam-constrained $B$ meson mass
$M_{\rm bc} = \sqrt{E_{\rm beam}^2 - (\sum p_i)^2}$, where $E_{\rm beam}$ 
is the CM beam 
energy, and $E_i$ and $p_i$ are the CM energies and momenta of the
$B$ candidate decay products. We select events with $M_{\rm bc}>5.2$ GeV/c$^2$
and $|\Delta E|<0.2$ GeV for the analysis. The requirements for signal 
candidates are $5.272$~GeV/$c^2<M_{\rm bc}<5.288$ GeV/$c^2$ 
and $|\Delta E|<0.022$ GeV. 
In addition, we make a  requirement on the invariant mass of the 
neutral $D$ candidate: 
$|M_{K_S\pi\pi}-M_{D^0}|<11$ MeV/$c^2$. 


To suppress background from $e^+e^-\to q\bar{q}$ ($q=u, d, s, c$) 
continuum events, we require $|\cos\theta_{\rm thr}|<0.8$, 
where $\theta_{\rm thr}$ is the angle between the thrust axis of 
the $B$ candidate daughters and that of the rest of the event. 
For additional background rejection, we 
use a Fisher discriminant composed of 11 parameters \cite{fisher}: 
the production angle of the $B$ candidate, the angle of the $B$ thrust 
axis relative to the beam axis and nine parameters representing 
the momentum flow in the event relative to the $B$ thrust axis in the CM frame.
We apply a requirement on the Fisher 
discriminant that retains 90\% of the signal and rejects 40\% of the 
remaining continuum background. 

\begin{figure}
  \epsfig{figure=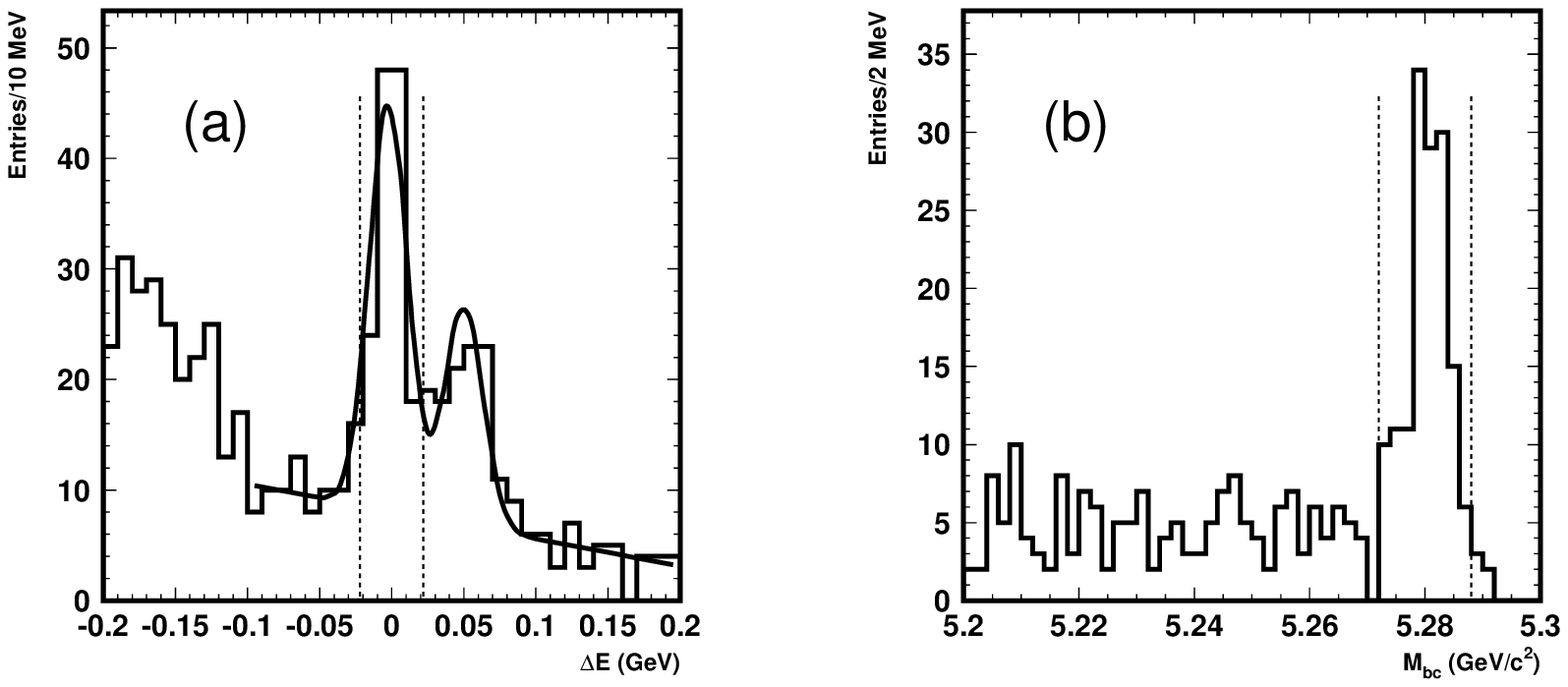,width=\textwidth}
  \caption{(a) $\Delta E$ and (b) $M_{\rm bc}$ distributions for the \bdk
  candidates. Dashed lines show the signal region. 
  The histogram shows the data; the smooth curve in (a) is the fit result.}
  \label{b2dk_sel}
\end{figure}

The $\Delta E$ and $M_{\rm bc}$ distributions for \bdk\ candidates are
shown in Fig.~\ref{b2dk_sel}. The peak in the $\Delta E$ distribution at 
$\Delta E=50$ MeV is due to \bdpi\ decays, where the pion is misidentified
as a kaon.
The ratio of the number of events in the peak at $\Delta E=50$ MeV
and in the signal peak is $0.54\pm 0.11$, which is consistent with 
the ratio of \bdk\ and \bdpi\ branching fractions of 
$0.079\pm 0.009\pm 0.006$ \cite{bdk_belle} and a 5\% $\pi/K$ misidentification
probability for our $\mathcal{R}_{\rm PID}(K)$ requirement. 
The \bdk\ selection efficiency (11\%) is determined from 
a Monte Carlo (MC) simulation. The number of events passing all selection 
criteria is 146. 
The background fraction is determined from a binned fit to the $\Delta E$
distribution, in which the signal is represented by a Gaussian distribution 
with mean
$\Delta E=0$, the \bdpi\ component is represented by a Gaussian 
distribution with 
mean $\Delta E=50$ MeV and the remaining background is modeled by a 
linear function. The contributions in the signal region are found 
to be $112\pm 12$ signal events, $1.1\pm 0.2$ \bdpi\ events and 
$35\pm 3$ events in the linear background. 
The overall background fraction is $25\pm 4$\%.

\subsection{Selection of \boldmath{\bdsk}}

\begin{figure}
  \epsfig{figure=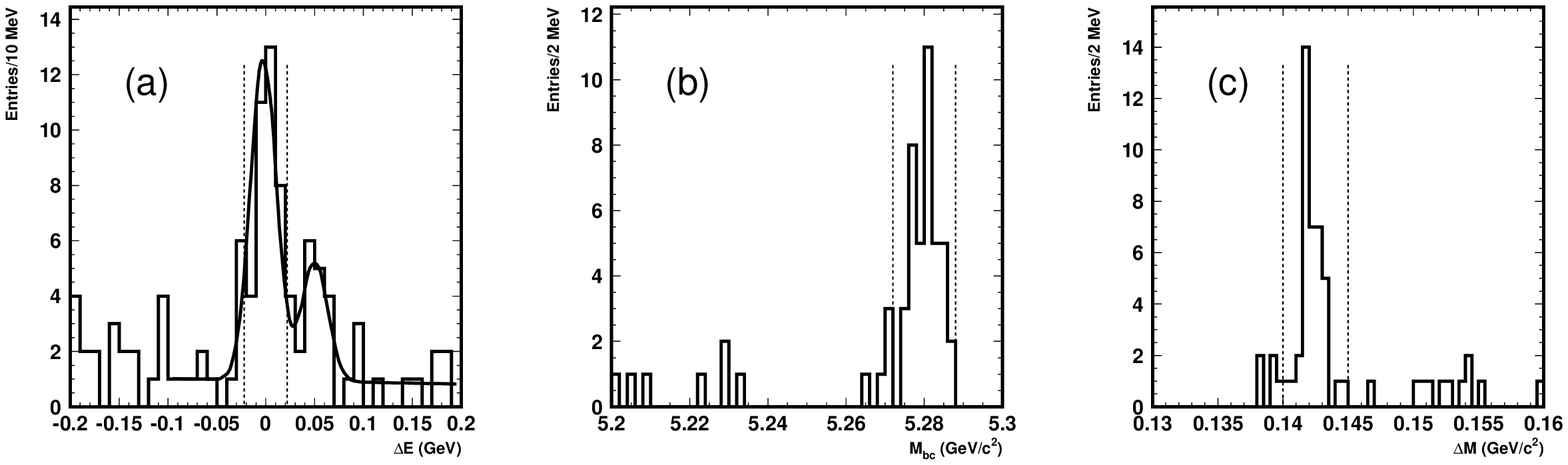,width=\textwidth}
  \caption{(a) $\Delta E$, (b) $M_{\rm bc}$ and (c) $\Delta M$ 
           distributions for the \bdsk candidates. 
           Dashed lines show the signal region. The histogram shows the data;
           the smooth curve in (a) is the fit result.}
  \label{b2dsk_sel}
\end{figure}

For the selection of \bdsk\ events, in addition to the 
requirements described above, we require the mass difference 
$\Delta M=M_{K_S\pi^+\pi^-\pi^0}-M_{K_S\pi^+\pi^-}$ of 
neutral $D^{*}$ and $D$ candidates to satisfy 
$140\mbox{ MeV}/c^2<\Delta M<145\mbox{ MeV}/c^2$.
Figure \ref{b2dsk_sel} shows the $\Delta E$, $M_{\rm bc}$ and $\Delta M$ 
distributions for \bdsk\ candidates. The selection 
efficiency is 6.2\%. The number of events satisfying 
the selection criteria is 39. The background fraction is determined in 
the same way as for \bdk\ events. The fit of the $\Delta E$
distribution yields $34\pm 6$ signal events, $4.4\pm 1.1$ events 
corresponding to the linear background and $0.24\pm 0.08$ \bdspi\ 
events in the signal region. The background fraction is $12\pm 4$\%.

\section{Determination of \boldmath{\dkpp} decay model}

The amplitude $f$ of the \dkpp\ decay is represented
by a coherent sum of two-body decay amplitudes plus one non-resonant 
decay amplitude,
\begin{equation}
  f(m^2_+, m^2_-) = \sum\limits_{j=1}^{N} a_j e^{i\alpha_j}
  \mathcal{A}_j(m^2_+, m^2_-)+
    b e^{i\beta}, 
  \label{d0_model}
\end{equation}
where $N$ is the total number of resonances, 
$\mathcal{A}_j(m^2_+, m^2_-)$, $a_j$ and 
$\alpha_j$ are the matrix element, amplitude and phase, respectively, 
of the $j$-th resonance, and $b$ and $\beta$ are the amplitude
and phase of the non-resonant component. The total phase and amplitude 
are arbitrary. To be consistent with a CLEO analysis \cite{dkpp_cleo}, 
we have chosen the $\bar{D^0}\to K_S\rho$ 
mode to have unit amplitude and zero relative phase. 
The description of the matrix elements follows Ref.~\cite{cleo_model}. 
The matrix elements for the resonances are parameterized by
Breit-Wigner shapes with $D$ meson and intermediate resonance 
form factors and angular dependences taken into account. 
If we consider the decay of $\bar{D^0}$ into a particle $C$ and a 
resonance $r$, with spin $J$, that subsequently decays into 
particles $A$ and $B$, the expression for the matrix element is
\[
  \mathcal{A}=F_D F_r\frac{s_J}{M_r^2-M^2_{AB}-iM_r\Gamma_{AB}},
\]
where $M_r$ is the mass of the resonance, $M_{AB}$ is the invariant mass
of the $AB$ system, $F_D$ and $F_r$ are the form factors of the $\bar{D^0}$
and the resonance, respectively, $\Gamma_{AB}$ is the mass dependent width 
of the resonance, and $s_J$ accounts for the angular momentum of the 
resonance. The form factors are the Blatt-Weisskopf 
penetration factors \cite{formfactors}; both depend on the spin $J$ of the 
intermediate resonance. We use the following expressions for the form 
factors:
\[
  F=1
\]
for $J=0$, 
\[
  F=\sqrt{\frac{1+R^2 p^2_r}{1+R^2 p^2_{AB}}}
\]
for $J=1$, and 
\[
  F=\sqrt{\frac{9+3R^2 p^2_r+R^4 p^4_r}{9+3R^2 p^2_{AB}+R^4 p^4_{AB}}}
\]
for $J=2$. Here $R$ is a parameter that describes the radial size of the meson
(either $\bar{D^0}$ or resonance $r$), 
$p_{AB}$ is the daughter particle momentum in the meson rest frame
and $p_r$ is its value when $M_{AB}=M_r$
(for $F_D$ the daughters are $C$ and the resonance $r$, 
for $F_r$ the daughters are $A$ and $B$). 
The radial parameters we use are 
$R=5$ GeV$^{-1}$ for the $\bar{D^0}$ and $R=1.5$ GeV$^{-1}$ for 
all intermediate resonances. The mass dependent width is given by
\[
  \Gamma_{AB}=\Gamma_r\left(\frac{p_{AB}}{p_r}\right)^{2J+1}
                      \left(\frac{M_r}{M_{AB}}\right)F_r^2, 
\]
where $\Gamma_r$ is the width of the resonance. 
The angular term $s_J$ depends on the spin of the resonance. The expressions
for scalar, vector and tensor states are:
\[
  s_0 = 1, 
\]
\[
  s_1 = M^2_{AC}-M^2_{BC}+\frac{(M^2_D-M^2_C)(M^2_B-M^2_A)}{M_r^2}, 
\]
\begin{eqnarray*}
  s_2 &=&
  \left(M^2_{BC}-M^2_{AC}+\frac{(M^2_D-M^2_C)(M^2_B-M^2_A)}{M_r^2}\right)^2-\\
  & &\frac{1}{3}\left( M^2_{AB}-2M^2_D-2M^2_C+\frac{(M^2_D-M^2_C)^2}{M_r^2}\right)
                \left( M^2_{AB}-2M^2_A-2M^2_B+\frac{(M^2_A-M^2_B)^2}{M_r^2}\right).
\end{eqnarray*}


For the $\bar{D^0}$ model we use a set of 15 two-body amplitudes. 
These include four Cabibbo-allowed amplitudes: $K^*(892)^+\pi^-$, $K_0^*(1430)^+\pi^-$, 
$K_2^*(1430)^+\pi^-$ and $K^*(1680)^+\pi^-$;  
doubly Cabibbo-suppressed partners for each of these states; and seven
channels with $K_S$ and a $\pi\pi$ resonance:
$K_S\rho$, $K_S\omega$, $K_Sf_0(980)$, $K_Sf_2(1270)$, 
$K_Sf_0(1370)$, $K_S\sigma_1$ and $K_S\sigma_2$. The masses and Breit-Wigner 
widths of the scalars $\sigma_1$ and $\sigma_2$ are left unconstrained, while 
the masses and widths of other resonances are taken to be the same as 
in the CLEO analysis \cite{dkpp_cleo}. In contrast to the CLEO analysis, we 
have introduced all doubly Cabibbo-suppressed amplitudes for flavor-specific 
decays (only $K^*(892)^-\pi^+$ was considered by CLEO) and two scalar
states $\sigma_1$ and $\sigma_2$. The amplitude for $\sigma_1$ describes 
the excess of 
events near the low $\pi\pi$ invariant mass edge of the phase space. 
The resonance $\sigma_2$ was 
introduced to describe a structure near 1.1 GeV$^2/c^4$ in the $m^2_{\pi\pi}$
distribution. This structure could be due to the 
decay $f_0(980)\to\eta\eta$ with rescattering of 
$\eta\eta$ to $\pi^+\pi^-$, which could distort 
the $f_0(980)\to \pi^+\pi^-$ amplitude for $m_{\pi\pi}$ 
near the $\eta\eta$ production threshold. 


We use an unbinned maximum likelihood technique to fit the Dalitz plot 
distribution to the model described by Eq.~\ref{d0_model}. 
We minimize the inverse 
logarithm of the likelihood function in the form
\begin{equation}
  -2 \log L = -2\left[\sum\limits^n_{i=1}\log p(m^2_{+,i}, m^2_{-,i}) - 
  \log\int\limits_D p(m^2_+, m^2_-)dm^2_+ d m^2_-\right], 
  \label{log_l}
\end{equation}
where $i$ runs over all selected event candidates, and
$m^2_{+,i}$, $m^2_{-,i}$ are measured Dalitz plot
variables. The integral in the second term accounts for the overall 
normalization of the probability density. 

The Dalitz plot density is represented by
\begin{equation}
  p(m^2_+, m^2_-) = \varepsilon(m^2_+, m^2_-)
  \int\limits^{\infty}_{-\infty}|M(m^2_++\mu^2, m^2_-+\mu^2)|^2
  \exp\left(-\frac{\mu^2}{2\sigma^2_{m}(m^2_{\pi\pi})}\right)d\mu^2
  +B(m^2_+, m^2_-),
  \label{density}
\end{equation}
where $M(m^2_+, m^2_-)=f(m^2_+, m^2_-)$ is the decay amplitude described 
by Eq.~\ref{d0_model}, 
$\varepsilon(m^2_+, m^2_-)$ is the efficiency, 
$B(m^2_+, m^2_-)$ is the background density, 
$\sigma_m(m^2_{\pi\pi})$ is the resolution of the squared invariant 
mass $m^2_{\pi\pi}$ of two pions
($m^2_{\pi\pi}=M^2_D+M^2_K+2M^2_{\pi}-m^2_+-m^2_-$). 
The free parameters of the minimization are the amplitudes
$a_j$ and phases $\alpha_j$ of the resonances (except for the $K_S\rho$
component, for which the parameters are fixed), 
the amplitude $b$ and phase $\beta$ of the non-resonant component
and the masses and widths of the $\sigma_1$ and $\sigma_2$ scalars. 

The background density for \dkpp\ events is extracted from 
$\Delta M$ sidebands: $\Delta M<142$ MeV$/c^2$ and 
$148$ MeV$/c^2<\Delta M<150$ MeV$/c^2$. 
The background density is parameterized by a third-order polynomial 
in the variables $m^2_+$ and $m^2_-$ to describe the
purely combinatorial background, plus Dalitz plot densities for
$D^0$ and $\bar{D^0}$ decays that correspond
to events where a correctly reconstructed $\bar{D^0}$ is combined with a 
random slow pion. 
From the fit, we obtain the fractions of the background components:
the purely combinatorial background is $43\pm 3$\%, 
combinations of $D^0$ with a pion of the correct charge account
for $49\pm 3$\%, and the remaining $8\pm 1$\% is due to
$D^0$'s combined with a pion of the wrong charge. 
The background fraction is fixed to 3.1\% from the fit 
to the $\Delta M$ distribution. As a consistency check, we also 
perform a fit with the background fraction floated and obtain a value for the 
background fraction in agreement with the result from the $\Delta M$ fit.

Our analysis is not sensitive to the absolute value of the 
reconstruction efficiency, but variations of the efficiency 
across the phase space can effect the fit result. 
The shape of the efficiency over the Dalitz plot is extracted from a
MC simulation, where the $\bar{D^0}$ decays uniformly over the allowed
phase space. The parameterization of the efficiency shape is a 
third-order polynomial in the variables $m^2_+$ and $m^2_-$, and
symmetrical under interchange of $\pi^+$ and $\pi^-$.
The efficiency is nearly uniform over the central part of the Dalitz plot, 
but drops by between 5\% and 13\% at the edges of phase space. 
Finite momentum resolution has to be taken into account in 
the fit function since our model includes a narrow 
$\omega\to \pi^+\pi^-$ state. 
Although both $m^2_+$ and $m^2_-$ have finite resolution, 
we consider only the resolution of the $m^2_{\pi\pi}$ combination, 
since the other Dalitz plot projections do not 
contain any narrow structures.
The resolution $\sigma_m$ as a function of $m_{\pi\pi}$ is 
parameterized by a linear function and is extracted from 
MC simulation. The average $m_{\pi\pi}^2$ resolution is 
$4.8\times 10^{-3}$~GeV$^2/c^4$. 

\begin{figure}	
  \vspace{-0.05\textwidth}
  \begin{center}
  \epsfig{figure=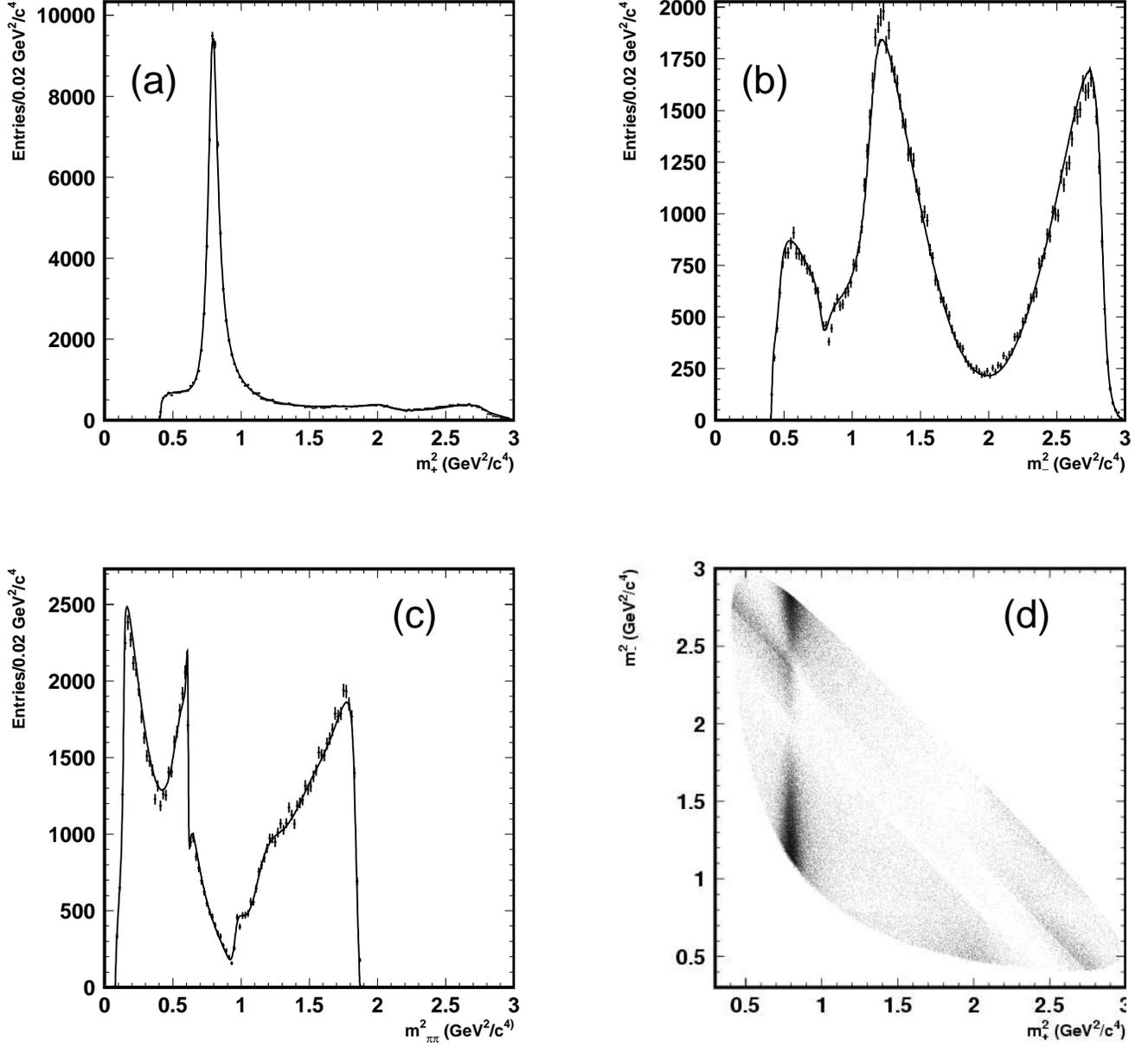,width=\textwidth}
  \caption{(a) $m^2_+$, (b) $m^2_-$, (c) $m^2_{\pi\pi}$
   distributions
   and (d) Dalitz plot for the \dkpp\ decay from the \dsdpis\ process.
   The points with error bars show the data, the smooth curve is the 
   fit result.}
  \label{ds2dpi_plot}
  \end{center}
\end{figure}

The fit results are given in Table~\ref{dkpp_table}. 
The parameters of the $\sigma$ resonances obtained 
in the fit are: 
$M_{\sigma_1}=539\pm 9$ MeV/$c^2$,
$\Gamma_{\sigma_1}=453\pm 16$ MeV/$c^2$, 
$M_{\sigma_2}=1048\pm 7$ MeV/$c^2$, and
$\Gamma_{\sigma_2}=109\pm 11$ MeV/$c^2$.
The \dkpp\ Dalitz plot, as well as
its projections with the fit results superimposed, are shown in 
Fig.~\ref{ds2dpi_plot}. The large peak in the $m^2_+$ distribution 
corresponds to the dominant $\bar{D^0}\to K^*(892)^+\pi^-$ mode. 
The minimum in the $m^2_-$ distribution at 0.8~GeV$^2/c^4$
is due to destructive interference with the doubly Cabibbo 
suppressed $\bar{D^0}\to K^*(892)^-\pi^+$ amplitude. In the $m^2_{\pi\pi}$
distribution, the $\bar{D^0}\to K_S\rho$ contribution 
is visible around 0.5~GeV$^2/c^4$
with a steep edge on the upper side due to interference with 
$\bar{D^0}\to K_S\omega$. The minimum around 0.9~GeV$^2/c^4$ is due to 
the decay $\bar{D^0}\to K_S f_0(980)$ interfering destructively with
other modes.

We obtain a larger amplitude for the non-resonant component compared to the 
CLEO analysis \cite{dkpp_cleo} (the fit fraction corresponding to the 
non-resonant component in our case is 24\%). The non-resonant component
is found to be highly correlated with the 
amplitude for the $\sigma_1$ resonance. A fit with the non-resonant 
amplitude fixed to zero yields a $\sigma_1$ amplitude of $0.78\pm 0.05$, 
while a fit without the $\sigma_1$ yields a non-resonant amplitude
of $4.66\pm 0.15$. Therefore, we conclude that the large non-resonant fraction 
in our $\bar{D^0}$ model is due to a deficiency in our description of 
the $\sigma_1$ state. We include this effect in the model uncertainty by
performing additional fits to the \bddsk\ data with $\bar{D^0}$ models 
with the non-resonant or $\sigma_1$ amplitudes excluded. 

The unbinned likelihood technique does not provide a reliable 
criterion for the goodness of fit. To check the quality of the 
fit, we make use of the large number of events in our sample and
perform a binned $\chi^2$ test by dividing the Dalitz plot into 
square regions $0.05\times 0.05$ GeV$^2/c^4$. The test yields 
$\chi^2=2121$ for 1130 degrees of freedom. More detailed studies 
are required in order to understand the precise dynamics of 
\dkpp\ decay. However, for the purpose of measuring $\phi_3$, 
we take the fit discrepancy into account in the model uncertainty. 


\section{Dalitz plot analysis of \boldmath{\bdk} decay}

\begin{figure}
  \epsfig{figure=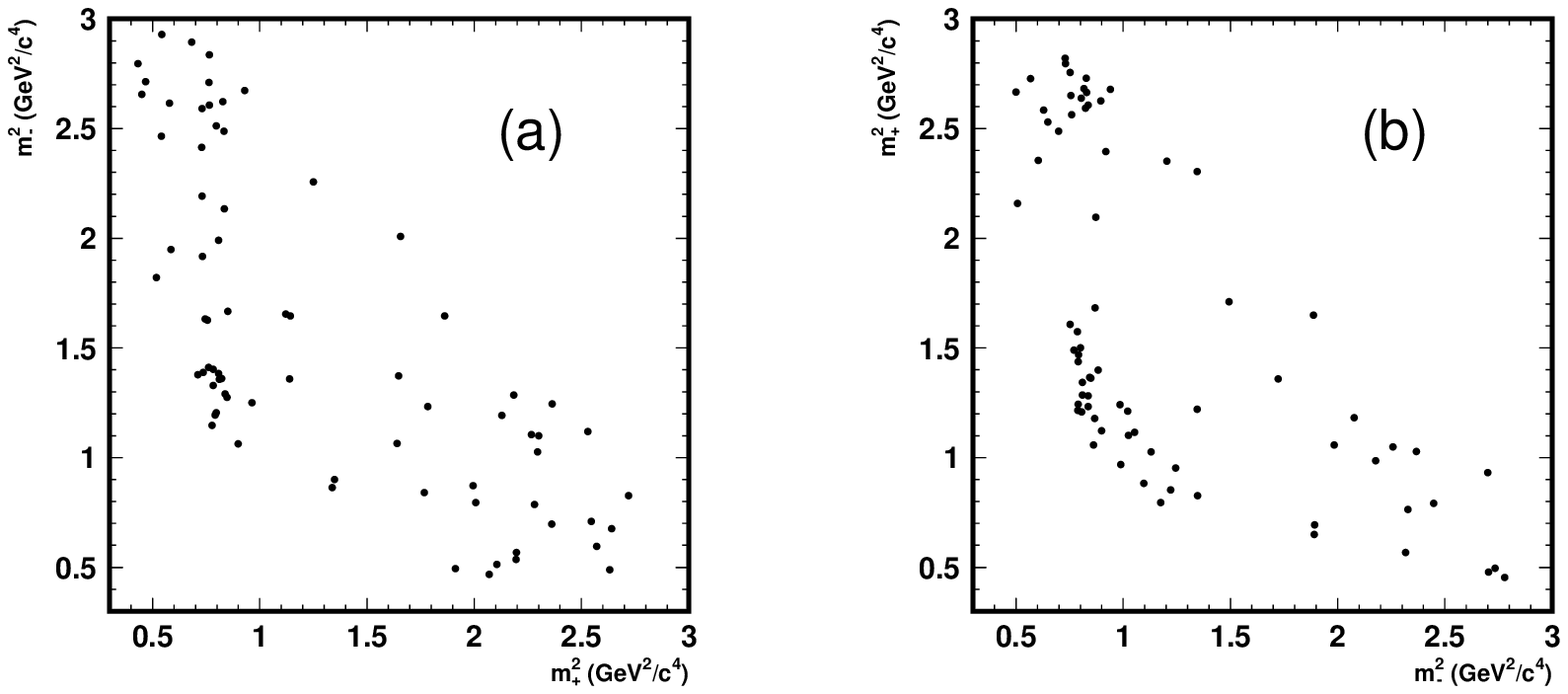,width=\textwidth}
  \caption{Dalitz plots of \dtkpp\ decay from (a) $B^+\to \tilde{D} K^+$
  and (b) $B^-\to \tilde{D} K^-$.}
  \label{b2dk_plots}
\end{figure}

\begin{figure}
  \epsfig{figure=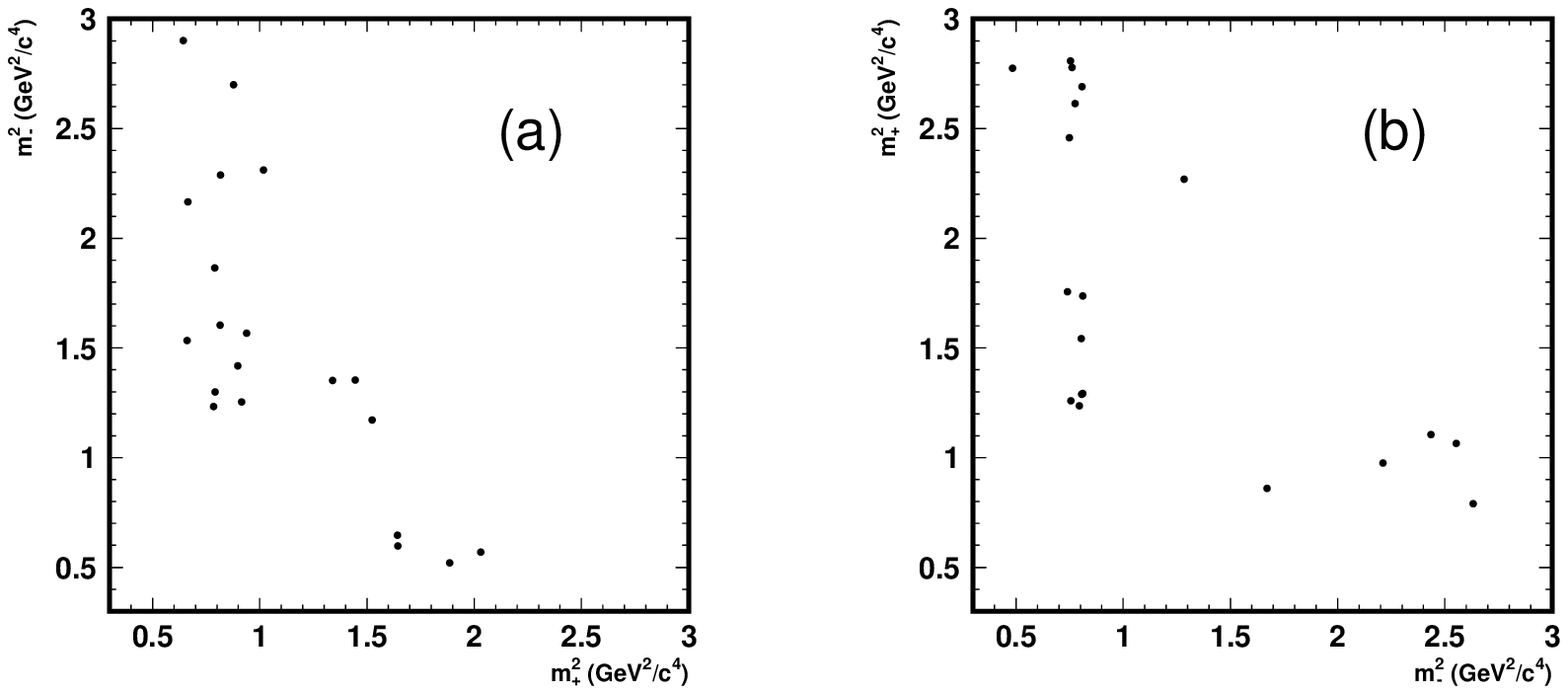,width=\textwidth}
  \caption{Dalitz plots of \dtkpp\ decay from (a) $B^+\to \tilde{D^{*}} K^+$
  and (b) $B^-\to \tilde{D^{*}} K^-$.}
  \label{b2dsk_plots}
\end{figure}

The Dalitz plots for \dtkpp, 
which contain information about $CP$ violation in $B$
decays, are shown in Figs.~\ref{b2dk_plots} and \ref{b2dsk_plots}
for \bdtk\ and \bdstk, respectively. These distributions are 
fitted by minimizing the combined logarithmic 
likelihood function 
\[
  -2\log L=-2\log L_- -2\log L_+,
\]
where $L_-(L_+)$ are the likelihoods of $B^-(B^+)$ data given 
by Eq.~\ref{log_l}. The corresponding Dalitz plot densities 
$p_{\pm}(m_+^2, m_-^2)$ are given by Eq.~\ref{density} with 
decay amplitudes $M_{\pm}$ described by Eq.~\ref{intdist} ($B^+$ data)
and Eq.~\ref{intdist_m} ($B^-$ data). 
The $\bar{D^0}$ decay model $f$ is fixed, and the free parameters of the
fit are the amplitude ratio $r$ and phases $\phi_3$ and $\delta$. 

As in the study of the sample from continuum \dsdpi\ decays, 
the efficiency and the momentum 
resolution were extracted from the signal MC sample, where the neutral $D$
meson decays according to phase space. The determination of the background 
contribution is described below. 

\subsection{Backgrounds}


Five sources of background are considered in our analysis 
(see Table~\ref{bck_table}). We determine the fraction and 
Dalitz plot shape for each component and use the results in the fit
to the signal Dalitz plot.
The largest contribution comes from two kinds of continuum events:
random combination of tracks, and correctly
reconstructed neutral $D$ mesons combined with random kaons.
These backgrounds are analyzed using an off-resonance
data sample collected at an energy 60 MeV below the $\Upsilon(4S)$ resonance 
in addition to a sample in which we make requirements 
on $|\cos\theta_{\rm thr}|$ and the Fisher discriminant that
select continuum events and reject $B\bar{B}$ events. 
The continuum background fraction is 
$22.1\pm 3.9$\% for \bdk\ and $9.0\pm 3.6$\% for \bdsk. The Dalitz 
plot shape of the continuum background is parameterized by a 
third-order polynomial in the variables $m^2_+$ and $m^2_-$ for the 
combinatorial background component and a sum of $D^0$ and $\bar{D^0}$ shapes for real 
neutral $D$ mesons combined with random kaons. 
 
The background from $B\bar{B}$ events originates either from a
\bddsk\ decay with some of the final state particles replaced by the 
decay products of the other $B$ meson, or from other
charged or neutral $B$ decays (possibly with misidentified or lost 
particles). We subdivide the $B\bar{B}$ background into four categories:
\begin{enumerate}
  \item 
Decays other than \bddsk\ and \bddspi, of which the dominant fraction
comes from the decay of $D^{(*)}$ from one $B$ meson, 
with some particles taken from the other $B$ decay, 
constitute the largest part of the $B\bar{B}$ background. 
They are investigated with a generic MC sample. For this, we obtain background
fractions of $2.2\pm 0.2$\% for \bdk\ and $2.1\pm 0.4$\%
for \bdsk. The parameterization of this background includes a linear 
function in the variables $m^2_+$ and $m^2_-$ and a Gaussian peak in $m^2_-$. 

  \item 
The process \bddspi\ with a pion misidentified as a kaon is suppressed by the
requirement on the $K/\pi$ identification probability and on the 
CM energy difference, $\Delta E$. 
The fraction of this background is obtained by fitting the $\Delta E$ 
distribution; the corresponding Dalitz plot distribution is that of 
$\bar{D^0}$ without the opposite flavor admixture. 
The fractions for this background are $1.0\pm 0.2$\% 
for \bdk\ and $0.6\pm 0.2$\% for \bdsk.

  \item 
\bddsk\ events where one of the neutral $D$ meson decay products is 
replaced by
a random kaon or pion were studied using a MC data set where 
one of the charged $B$ mesons from the $\Upsilon(4S)$ decays into the 
$D^{(*)}K$ state. 
The corresponding background fraction is $0.4\pm 0.1$\% 
for both \bdk\ and \bdsk\ modes; the Dalitz plot shape is parameterized
by a linear function in the variables $m^2_+$ and $m^2_-$ 
plus a $D^0$ amplitude. 

  \item 
Events in which a correctly reconstructed neutral $D$ is combined 
with a random charged kaon 
are of importance, because half of the kaons have 
the wrong sign: such events will be misinterpreted 
as decays of $D$ mesons of the opposite flavor, and thus introduce distortion 
in the
most sensitive area of the Dalitz plot. In the MC sample, we find no events 
of this kind, which allows us to set an upper limit of 0.4\%
(at 95\% CL) on the fraction for this contribution. 

\end{enumerate}

\subsection{Control sample fits}

To test the consistency of the fitting procedure, the same fitting procedure was 
applied to the \bddstpi\ and $\bar{B^0}(B^0)\to D^{*\pm}\pi^{\mp}$ control samples 
as to the \bddstk\ signal. 
For decays in which only one flavor $D$ meson can contribute, the fit 
should return values of the amplitude ratio $r$ consistent with zero. 
In the case of \bddstpi\, a small amplitude ratio is expected 
($r\sim |V_{ub} V^*_{cd}|/|V_{cb}V^*_{ud}|\sim 0.02$). Deviations 
from these values can appear if the Dalitz plot distribution is not 
well described by the fit model. 

For the control sample fits, we consider $B^+$ and $B^-$ data separately, 
to check for the absence of $CP$ violation. 
The free parameters of the Dalitz plot fit are $r_{\pm}$ and 
$\theta_{\pm}$, where $\theta_{\pm}=\delta\pm\phi_3$ (see Eqs.~\ref{intdist}, 
\ref{intdist_m}). 

\begin{figure}
  \epsfig{figure=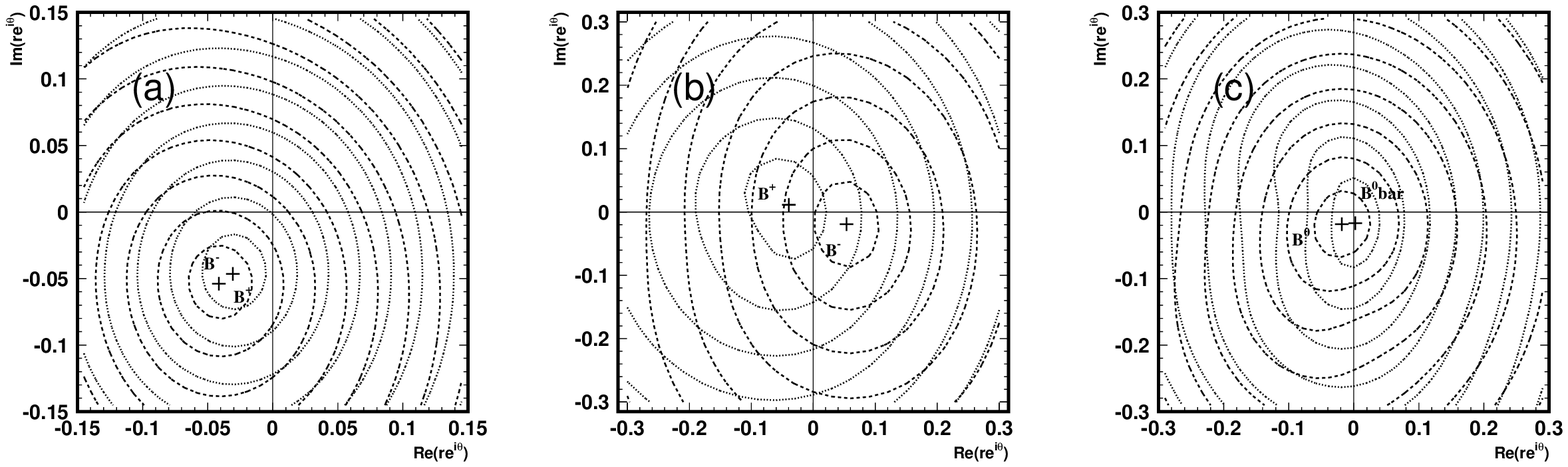,width=\textwidth}
  \caption{Constraint plots of the complex amplitude ratio
           $re^{i\theta}$ for (a) \bdtpi, 
           (b) \bdstpi\ and 
           (c) $\bar{B^0}(B^0)\to D^{*\pm}\pi^{\mp}$ decays.
	   Contours indicate integer multiples of the standard deviation.
           Dotted contours are from $B^+(\bar{B^0})$ data, dashed 
           contours are from $B^-(B^0)$ data.}
  \label{test_constr}
\end{figure}

The fit results for \bdtpi\ are $r_+=0.056\pm 0.028$, 
$\theta_+=237^{\circ}\pm 27^{\circ}$
for $B^+$ data and 
$r_-=0.068\pm 0.026$, 
$\theta_-=232^{\circ}\pm 22^{\circ}$ for $B^-$ data. 
It should be noted that since the value 
of $r$ is positive definite, the error of this parameter does not serve as a good 
measure of the $r=0$ hypothesis. To demonstrate the deviation of the 
amplitude ratio $r$ from zero, the real and imaginary parts of the complex 
amplitude ratio $r e^{i\theta}$ are more suitable. 
Figure~\ref{test_constr} (a) shows the complex 
amplitude ratio constraints for the $B^+$ and $B^-$ data separately. 
It can be seen from the plot that both amplitude ratios differ from 
the expected 
value by more than two standard deviations. This deviation is treated as 
a potential systematic effect. 

The other control samples, \bdstpi\ with $\tilde{D}^{*}$ decaying to 
$\tilde{D}\pi^0$, and $\bar{B^0}(B^0)\to D^{*\pm}\pi^{\mp}$ with 
$D^{*\pm}\to D\pi^\pm$, 
do not show any significant deviation from $r=0$. 
The results of the fit to the \bdstpi\ sample (351 events) are 
$r_+=0.041\pm 0.069$, $\theta_+=163^{\circ}\pm 100^{\circ}$, 
$r_-=0.057\pm 0.054$, $\theta_-=340^{\circ}\pm 65^{\circ}$ and 
are shown in Fig.~\ref{test_constr} (b);
the results of the fit to the $\bar{B^0}(B^0)\to D^{*\pm}\pi^{\mp}$ 
sample (517 events) are
$r_+=0.017\pm 0.070$, $\theta_+=278^{\circ}\pm 133^{\circ}$, 
$r_-=0.026\pm 0.050$, $\theta_-=225^{\circ}\pm 99^{\circ}$
and are shown in Fig.~\ref{test_constr} (c).

\section{Results}

\begin{figure}
  \vspace{-2\baselineskip}
  \epsfig{figure=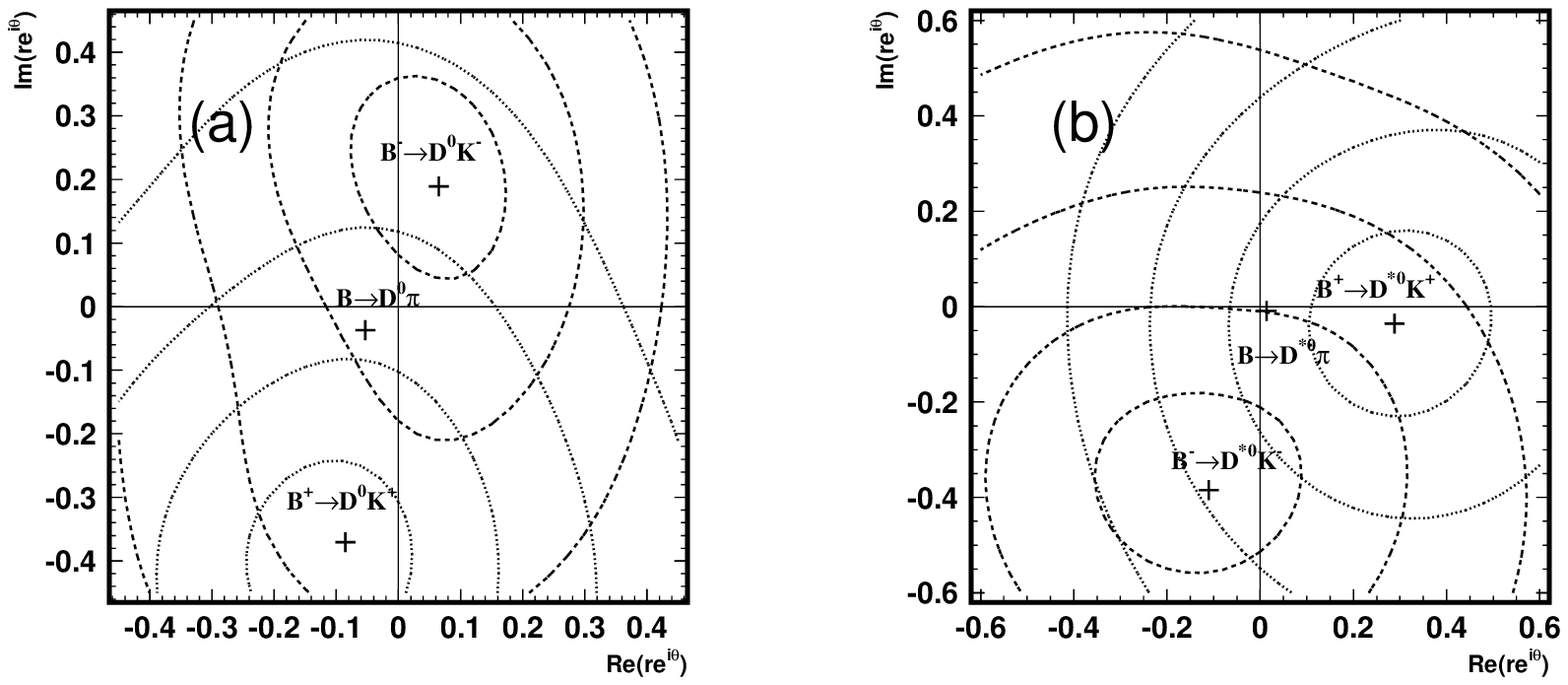,width=\textwidth}
  \caption{Constraint plots of the complex amplitude ratio
           $re^{i\theta}$ for (a) \bdtk\ and (b) \bdstk\ decays. 
	   Contours indicate integer multiples of the standard deviation.
           Dotted contours are from $B^+$ data, dashed 
           contours are from $B^-$ data.}
  \label{compl_constr}
\end{figure}

The results of the separate $B^+$ and $B^-$ data fits are shown in 
Fig.~\ref{compl_constr}. The plots show the constraints on the complex 
amplitude ratio $r e^{i\theta}$ for the \bdtk\ and \bdstk\ samples. The 
fit technique is the same as the one used for the control 
samples. It can be seen that in both signal samples a significant
non-zero value of $r$ is observed.
A difference between the phases $\theta_+$ and $\theta_-$ 
is also apparent in both the \bdtk\ and \bdstk\ samples, 
which indicates a deviation of $\phi_3$ from zero. 

A combined unbinned maximum likelihood fit to the 
$B^+$ and $B^-$ samples with 
$r$, $\phi_3$ and $\delta$ as free parameters yields the following values: 
$r=0.31\pm 0.11$, $\phi_3=86^{\circ}\pm 17^{\circ}$, 
$\delta=168^{\circ}\pm 17^{\circ}$ for the \bdtk\ sample and 
$r=0.34\pm 0.14$, $\phi_3=51^{\circ}\pm 25^{\circ}$, 
$\delta=302^{\circ}\pm 25^{\circ}$ for the \bdstk\ sample.
The errors quoted here are obtained from the likelihood fit.
These errors are a good representation of the uncertainties for
a Gaussian likelihood distribution, however in our case
the distributions are highly non-Gaussian. In addition, the errors
for the strong and weak phases depend on the values of the
amplitude ratio $r$ ({\it e.g.} for $r=0$ there is 
no sensitivity to the phases). A more reliable estimate of the
statistical uncertainties is obtained using a large number
of MC pseudo-experiments as discussed below.

\subsection{Estimation of model uncertainty}

The model used for the \dkpp\ decay is one of the main sources of 
systematic error for our analysis. The model is a result of 
the fit to an experimental Dalitz plot, however, since the density 
of the plot is
proportional to the absolute value squared of the decay amplitude, 
the phase $\phi(m^2_+, m^2_-)$ of the complex amplitude is not 
directly measured. The phase variations across 
the Dalitz plot are therefore the result of model assumptions and 
their uncertainties may affect the $\tilde{D}$ Dalitz plot fit 
from \bddstk.

To estimate the effects of the model uncertainties, a MC simulation is used. 
Event samples are generated according to the Dalitz distribution 
described by the amplitude given by Eq.~\ref{intdist} 
with the resonance parameters extracted from 
our fit of continuum $D^0$ data, but to fit this simulated plot 
different models for $f(m_+, m_-)$ 
are used (see Table~\ref{model_table}). We scan the phases $\phi_3$
and $\delta$ in their physical regions and take the maximum 
deviations of the fit parameters ($(\Delta r)_{\rm max}$, 
$(\Delta\phi_3)_{\rm max}$, and $(\Delta\delta)_{\rm max}$) 
as model uncertainty estimates. 
The values for $(\Delta r)_{\rm max}$, 
$(\Delta\phi_3)_{\rm max}$ and $(\Delta\delta)_{\rm max}$
quoted in Table~\ref{model_table} are obtained with 
the value $r=0.13$. For larger $r$ values, the model uncertainty 
tends to be smaller, 
so our estimate of the model uncertainty is conservative. 

All the fit models are based on Breit-Wigner parameterizations 
of resonances as in our default model. Since a
Breit-Wigner amplitude can only describe narrow resonances well, 
the usual technique to deal with broad states is to introduce 
Blatt-Weisskopf form factors for the $\bar{D^0}$ meson ($F_D$) and intermediate 
resonance ($F_r$) and a $q^2$-dependence of the resonance width $\Gamma$. 
These quantities have substantial theoretical uncertainties 
and might introduce a large model error.
We have therefore used a fit model without Blatt-Weisskopf form factors 
and with a constant resonance width to estimate such an error.
We have also used a model containing only narrow resonances
($K^*(892)$, $\rho$, doubly Cabibbo-suppressed $K^*(892)$ and $f^0(980)$)
with the wide ones approximated by the flat non-resonant term.
The study of the model errors is summarized in Table~\ref{model_table}.
Our estimate of the systematic uncertainty on $\phi_3$ is $11^{\circ}$.


\subsection{Estimation of systematic errors}

In addition to the model uncertainty, there are other potential 
sources of systematic
error, such as uncertainties in the background Dalitz plot density, 
efficiency variations over the phase space and possible fit biases.
These are listed in Table~\ref{syst_table} for the \bdtk\ 
and \bdstk\ modes separately. The component related to the background 
shape parameterization was estimated by extracting the background shape 
from the $M_D$ sidebands and by using a flat background distribution.
The maximum deviation of the fit parameters from the ``standard"
background parameterization was assigned as the corresponding
systematic error. The effect of the uncertainty in the background 
fraction was studied by varying the background
fraction by one standard deviation.

A potentially dangerous background is caused by events 
with a random kaon, 
half of which would not have the correct charge. We set a 0.4\% 
upper limit on this kind of background at 95\% confidence level based on 
MC simulation. The effect of this background on the 
fit results was studied using a MC procedure similar to that used 
for investigating the model uncertainty. The bias in the fit parameters 
corresponding to a 0.4\% fraction is negligible ($0.7^{\circ}$ for $\phi_3$) 
in comparison to the systematic error due to background shape.

As mentioned above, the efficiency shape and momentum resolution 
were extracted from MC simulation. 
To estimate their contributions to the systematic error, we 
repeat the fit using a flat efficiency and a fit model that
does not take the resolution into account, respectively. 
The biases due to the efficiency shape differ for \bdtk\ and \bdstk\ 
samples, but since we expect the values of the efficiency systematics
to be close for the two modes, we assign the maximum value
of the bias as the corresponding systematic error. 

The non-zero amplitude ratio observed in the \bdtpi\ control sample
can be either due to a statistical fluctuation or 
may indicate some systematic effect such as background structure or a deficiency 
of the $\bar{D^0}$ decay model. Since the source of this bias is unknown, 
we conservatively treat it as an additional systematic effect. 
The corresponding bias of parameters is estimated in the following way. 
Suppose the parameters in the $re^{i\theta}$ plane are biased by a 
value $\Delta r$ in a certain direction, then the maximum bias of the 
total phases would be equal to 
$(\Delta\theta)_{max}\sim\Delta r/r\sim 11^{\circ}$. 
Since $\delta=(\theta_++\theta_-)/2$ and $\phi_3=(\theta_+-\theta_-)/2$, 
the maximum biases of the strong and weak phases would also be 
equal to $11^{\circ}$. 
The maximum bias of $r$ for the simultaneous fit to both flavors 
in the case of $\phi_3\sim \pi$ would be
$\Delta r=\sqrt{r^2+(\Delta r)^2}-r\sim 0.006$. 

\subsection{Evaluation of statistical error}

We use a frequentist technique to evaluate the 
statistical significance of the measurements. This method requires 
knowledge of the probability density function (PDF) of the 
fitted parameters as a function of the true parameters. 

To obtain this PDF, we employ a ``toy" MC technique that uses a
simplified MC simulation of the experiment which incorporates
the same efficiencies, resolution and backgrounds as
used in the fit to the experimental data.  This MC is used
to generate several hundred experiments for a given set of
$r$, $\theta_+$ and $\theta_-$ values. For each simulated
experiment, Dalitz plot distributions are generated
with numbers of events  that nearly equal to the numbers of
events observed in the data --- 70 events for each $B$ flavor for
\bdtk\ and 20 events for each $B$ flavor for
\bdstk\ ---  and the simulated Dalitz distributions
are subjected to the same fitting procedure that is applied
to the data. This is repeated for different values of $r$,
producing distributions of the fitted parameters that
are used to produce a functional form of the PDFs for 
reconstructed values for any set of input parameters.

We parameterize the PDF of a set of 
fitted parameters $(r,\phi_3,\delta)$, using the following model. 
We assume that
as a result of the fit of a single Dalitz plot (either $B^+$ or $B^-$ data), 
the errors of parameters 
$Re(r_{\pm} e^{i\theta_{\pm}})$ and $Im(r_{\pm} e^{i\theta_{\pm}})$ are 
uncorrelated and have Gaussian distributions with equal RMS
which we denote as $\sigma$. 
The PDF of the parameters $(r_{\pm}, \theta_{\pm})$ for the true 
parameters $(\bar{r}_{\pm}, \bar{\theta}_{\pm})$ is thus written as
\[
  d^2 P(r_{\pm},\theta_{\pm}| \bar{r}_{\pm},\bar{\theta}_{\pm})=
       \frac{1}{2\pi\sigma^2}\exp\left[-\frac{
       (r_{\pm}\cos\theta_{\pm}-\bar{r}\cos\bar{\theta}_{\pm})^2+
       (r_{\pm}\sin\theta_{\pm}-\bar{r}\sin\bar{\theta}_{\pm})^2
       }{2\sigma^2}\right] r_{\pm}dr_{\pm}d\theta_{\pm}. 
\]

To obtain the PDF for the parameters $(r,\phi_3,\delta)$ we fix
$r=r_+=r_-$ and substitute the total phases with $\delta+\phi_3$ and 
$\delta-\phi_3$:
\begin{equation}
  \frac{d^3 P}{dr d\phi_3 d\delta}(r,\phi_3,\delta| 
                       \bar{r}, \bar{\phi}_3, \bar{\delta})=
    \frac{d^2 P}{dr_+ d\theta_+}(r,\delta+\phi_3| \bar{r},
                                 \bar{\delta}+\bar{\phi}_3) 
    \frac{d^2 P}{dr_- d\theta_-}(r,\delta-\phi_3| \bar{r}, 
                                 \bar{\delta}-\bar{\phi}_3). 
  \label{fit_pdf}
\end{equation}
In this model, there is only one free parameter $\sigma$ which is obtained
from the unbinned maximum likelihood fit of the MC distribution of 
fitted parameters to Eq.~\ref{fit_pdf}. 
The values of $\sigma$ obtained for 
MC samples with different values of $r$ are all equal within
4\%, therefore we use a constant value $\sigma=0.14$ for the
PDF of \bdtk\ data. For the fit of the sample of neutral
$D$ mesons from \bdstk\ the value of $\sigma$ is 
equal to 0.22. 

After the PDF of the fitted parameters is obtained, the confidence 
level $\alpha$ for each set of true parameters 
$(\bar{r},\bar{\phi}_3,\bar{\delta})$ is defined as 
\[
  \alpha(\bar{r},\bar{\phi}_3,\bar{\delta}) = 
  \int_{\Omega}\frac{d^3 P}{dr d\phi_3 d\delta}(r, \phi_3, \delta|
           \bar{r},\bar{\phi}_3,\bar{\delta}) dr d\phi_3 d\delta, 
\]
where the corresponding confidence region $\Omega$ is given 
by the condition
\[
  \frac{d^3 P}{dr d\phi_3 d\delta}(r, \phi_3, \delta|
           \bar{r},\bar{\phi}_3,\bar{\delta})\geq
  \frac{d^3 P}{dr d\phi_3 d\delta}(0.31, 86^{\circ}, 168^{\circ}|
           \bar{r},\bar{\phi}_3,\bar{\delta}), 
\]
{\it i.e.} it includes all points in the fit parameter space 
for which the PDF is larger than that at the point 
$0.31, 86^{\circ}, 168^{\circ}$, corresponding to the fit result. 
For \bdstk\ these values are replaced by 0.34, $51^{\circ}$ and $302^{\circ}$
for $r$, $\phi_3$ and $\delta$, respectively. 

\begin{figure}
  \vspace{-2\baselineskip}
  \epsfig{figure=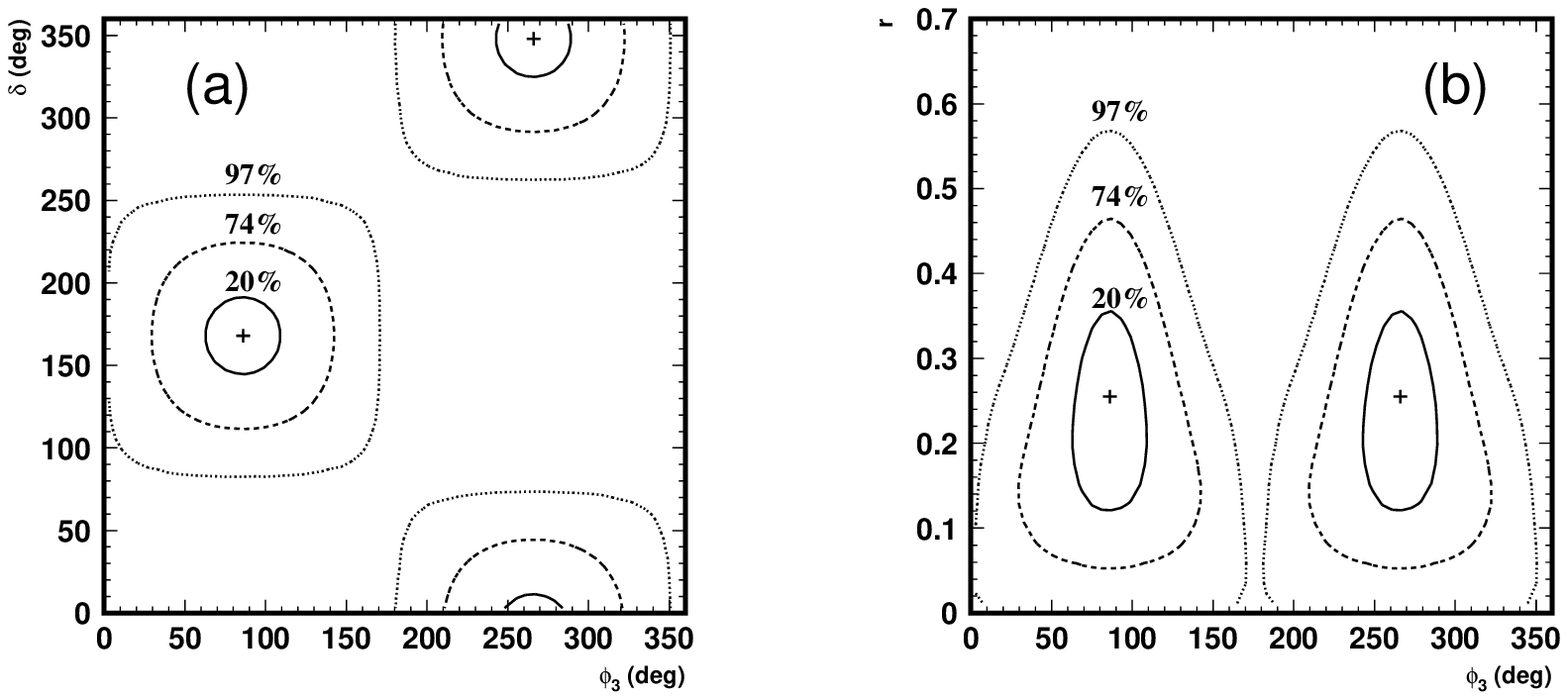,width=\textwidth}
  \caption{Confidence regions for the pairs of parameters (a) ($r$, $\phi_3$) 
           and (b) ($\phi_3, \delta$) for the \bdtk\ sample.}
  \label{b2dk_neum}
\end{figure}

\begin{figure}
  \vspace{-2\baselineskip}
  \epsfig{figure=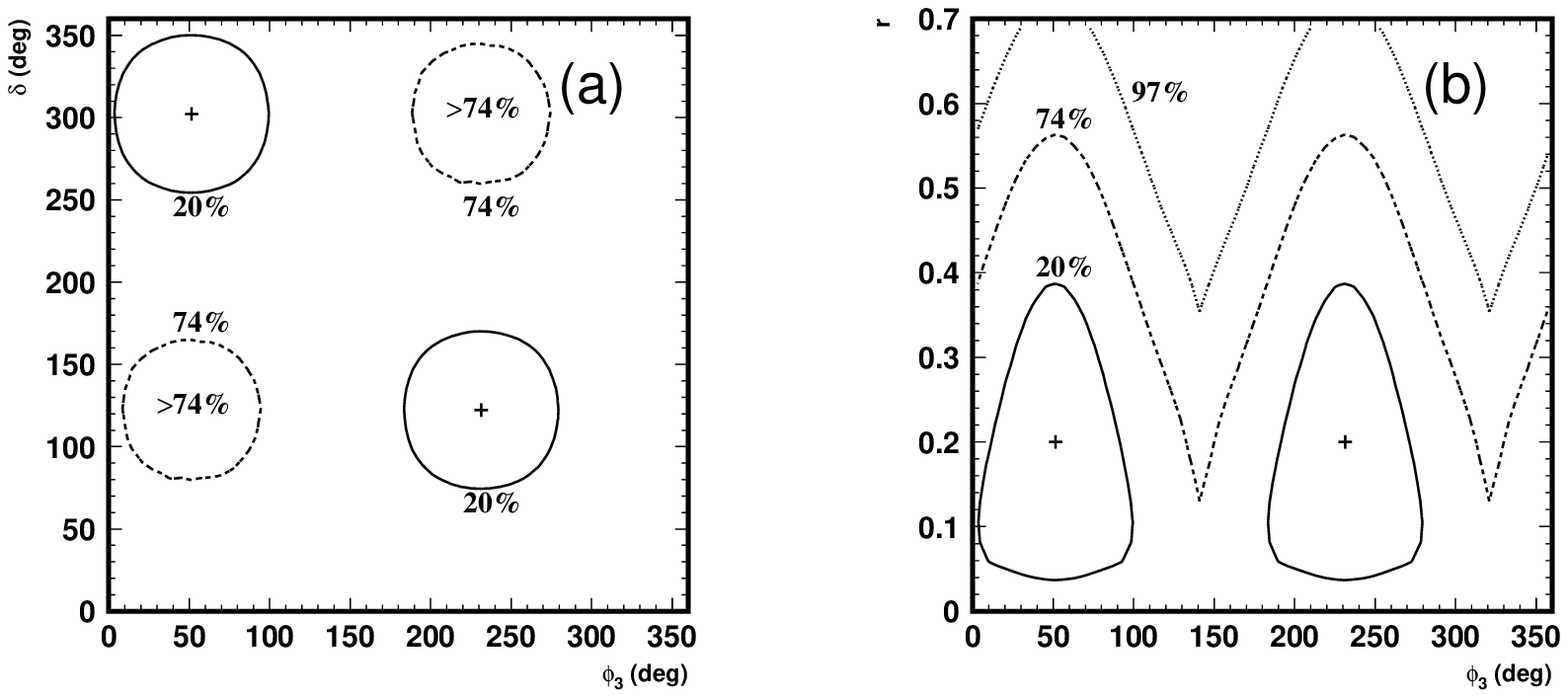,width=\textwidth}
  \caption{Confidence regions for the pairs of parameters (a) ($r$, $\phi_3$)
           and (b) ($\phi_3, \delta$) for the \bdstk\ sample.}
  \label{b2dsk_neum}
\end{figure}

The confidence regions for the pairs of parameters 
$(\phi_3, \delta)$ and $(\phi_3, r)$ are shown in Fig.~\ref{b2dk_neum} 
(\bdtk\ mode) and Fig.~\ref{b2dsk_neum} (\bdstk\ mode).
These plots are the projections of the corresponding 
confidence regions in the three-dimensional parameter space. 
We show the 20\%, 74\% and 97\% confidence level regions, 
which correspond to 
one, two, and three standard deviations for a three-dimensional Gaussian
distribution.
The 20\% confidence region, which corresponds to one standard deviation, 
yields the following results for the fit parameters: 
$r=0.26^{+0.10}_{-0.14}$, $\phi_3=86^{\circ}\pm 23^{\circ}$, 
$\delta=168^{\circ}\pm 23^{\circ}$ for \bdtk\ data and 
$r=0.20^{+0.19}_{-0.17}$, $\phi_3=51\pm 46^{\circ}$, $\delta=302\pm 46^{\circ}$
for \bdstk\ data. The central values presented are obtained by 
maximizing the fit parameters' PDF. This technique accounts for the 
parameter biases introduced by the fit procedure. We find that $\phi_3$ and
$\delta$ are unbiased in both the \bdk\ and \bdsk\ cases, while the 
central value of $r$ is biased by the fit procedure due to its positive 
definiteness.

The values of the amplitude ratio $r$ obtained are larger than 
our initial estimate ($r=0.13$), though they agree within the statistical
error. These values are also consistent with the recent 
measurement by BABAR collaboration \cite{babar}, which set up an
upper limit $r<0.22$ at 90\% CL for the \bdk\ mode. 

In the frequentist approach, the significance of the $CP$ violation 
is evaluated by finding the confidence level for the most probable 
$CP$ conserving point, {\it i.e.} the point with $r=0$ or $\phi_3=0$, 
for which the confidence level $\alpha(\bar{r},\bar{\phi}_3,\bar{\delta})$
is minimal. This procedure gives $\alpha=97$\% for the point 
$r=0.03$, $\phi_3=0$, $\delta=168^{\circ}$ for the \bdtk\ sample. 
The same procedure applied to 
\bdstk\ sample gives a $CP$ violation significance of 23\% 
(for the point $r=0.10$, $\phi_3=0$, $\delta=302^{\circ}$).

\subsection{Combined \boldmath{$\phi_3$} measurement using 
            \boldmath{\bdk} and \boldmath{\bdsk} samples}

The two events samples, \bdk\ and \bdsk, are combined 
in order to obtain a more accurate measurement of $\phi_3$. 
The technique we use to obtain the combined measurement is 
also based on a frequentist approach. 
Since in general the values of the amplitude ratio $r$ and 
strong phase $\delta$ can differ for the two modes, 
we have five true parameters ($\bar{\phi_3}$, $\bar{r}_1$, $\bar{r}_2$, 
$\bar{\delta}_1$ and $\bar{\delta}_2$, where the indices 1 and 2
correspond to \bdk\ and \bdsk\ modes, respectively) 
and six reconstructed parameters ($r$, $\phi_3$ and $\delta$ 
for each of the two modes). Since in this case the physical 
range of the parameters is smaller than the range of the reconstructed 
parameters ($\phi_3$ values have to be equal for the two modes), 
the Feldman-Cousins \cite{feldman} approach is used.

The PDF for the reconstructed parameters is written as
\[
  \frac{dP}{dx}(x,\mu)=
  \frac{d^3 P_{B\to D^0K}   }{dr d\phi_3 d\delta}
  (r_1,(\phi_3)_1,\delta_1|\bar{r_1},\bar{\phi}_3,\bar{\delta_1})
  \frac{d^3 P_{B\to D^{*0}K}}{dr d\phi_3 d\delta}
  (r_2,(\phi_3)_2,\delta_2|\bar{r_2},\bar{\phi}_3,\bar{\delta_2}), 
\]
where $x=(dr_1, d(\phi_3)_1, d\delta_1, dr_2, d(\phi_3)_2, d\delta_2)$ is a 
vector of the reconstructed parameters, and
$\mu=(\bar{\phi_3}, \bar{r}_1, \bar{r}_2, 
\bar{\delta}_1, \bar{\delta}_2)$ is a vector of the true parameters. 

The confidence level $\alpha$ for a vector of true parameters $\mu$
is defined as 
\[
  \alpha(\mu) = 
  \int_{\Omega}\frac{dP}{dx}(x|\mu) dx.
\]
The confidence region $\Omega$ is given by the Feldman-Cousins likelihood
ratio ordering:
\[
  \frac{dP}{dx}(x,\mu)\left/\frac{dP}{dx}(x,\mu_{best}(x))\right.>
  \frac{dP}{dx}(x_0,\mu)\left/\frac{dP}{dx}(x_0,\mu_{best}(x_0))\right.. 
\]
Here $\mu_{best}(x)$ is defined as the vector of true parameters that
maximizes the PDF for a given set $x$ of reconstructed parameters. 

The vector of the central values of the true parameters is given by
$\alpha=0$ and equals $\mu_{best}(x_0)$. The corresponding central 
value of $\phi_3$ is $77^{\circ}$.
The one standard deviation interval for $\phi_3$ 
(which corresponds to the 3.7\% confidence level for the case of 
a five-dimensional Gaussian distribution) is 
$\phi_3=77^{\circ}\;^{+17^{\circ}}_{-19^{\circ}}$;
the two standard deviation (or 45\% CL
for a five-dimensional distribution) interval is $39^{\circ}<\phi_3<112^{\circ}$. 
These intervals include only the statistical error. 
The statistical significance of $CP$ violation for the combined measurement 
is 95\%. 

Since the \bdk\ contribution dominates in the combined measurement, we 
use its value of the systematic uncertainty, which is $13^{\circ}$, as
an estimate of the systematic uncertainty in the combined $\phi_3$ 
measurement. The model uncertainty for the two modes is the same and amounts
to $11^{\circ}$. The two standard deviation interval including the 
systematic and model uncertainties is $26^{\circ}<\phi_3<126^{\circ}$. 

\section{Conclusion}

We report results of a measurement of the unitarity
triangle angle $\phi_3$ that uses a new method based on a 
Dalitz plot analysis of the three-body $D^0$ decay in the process 
\bddsk.
The first measurement of $\phi_3$ using this technique
was performed based on 140 fb$^{-1}$ data sample collected by 
the Belle detector. 
From the combination of \bdk\ and \bdsk\ modes, we obtain the 
value of 
$\phi_3=77^{\circ}\;^{+17^{\circ}}_{-19^{\circ}}\pm 13^{\circ}\pm 11^{\circ}$.
The first error is statistical, the second is experimental systematics and
the third is model uncertainty. 
The two standard deviation interval (including model and systematic 
uncertainties) is $26^{\circ}<\phi_3<126^{\circ}$.
The statistical significance of $CP$ violation for the combined 
measurement is 95\%. 
The method allows us to obtain a value of the $D^0$-$\bar{D^0}$
amplitude ratio $r$, which can be used in other $\phi_3$ 
measurements. We obtain $r=0.26^{+0.10}_{-0.14}\pm 0.03\pm 0.04$
for the \bdk\ mode and $r=0.20^{+0.19}_{-0.17}\pm 0.02\pm 0.04$ 
for the \bdsk\ mode.

The method has a number of advantages over the other
ways to measure $\phi_3$ \cite{glw}--\cite{ads}. 
It is directly sensitive to the value of $\phi_3$ and has only the two-fold 
discrete ambiguity ($\phi_3+\pi$, $\delta+\pi$). 
It does not involve branching 
fraction measurements and, therefore, the influence of the detector 
systematics is minimal. 
The statistical sensitivity of this technique is also superior
in the presence of background since an interference term is
measured. 

\section*{Acknowledgments}

We are grateful to V.~Chernyak and M.~Gronau for fruitful 
discussions. 
   We thank the KEKB group for the excellent
   operation of the accelerator, the KEK Cryogenics
   group for the efficient operation of the solenoid,
   and the KEK computer group and the National Institute of Informatics
   for valuable computing and Super-SINET network support.
   We acknowledge support from the Ministry of Education,
   Culture, Sports, Science, and Technology of Japan
   and the Japan Society for the Promotion of Science;
   the Australian Research Council
   and the Australian Department of Education, Science and Training;
   the National Science Foundation of China under contract No.~10175071;
   the Department of Science and Technology of India;
   the BK21 program of the Ministry of Education of Korea
   and the CHEP SRC program of the Korea Science and Engineering 
Foundation;
   the Polish State Committee for Scientific Research
   under contract No.~2P03B 01324;
   the Ministry of Science and Technology of the Russian Federation;
   the Ministry of Education, Science and Sport of the Republic of 
Slovenia;
   the National Science Council and the Ministry of Education of Taiwan;
   and the U.S.\ Department of Energy.

\newpage

\begin{table}
\caption{Fit results for \dkpp\ decay. Errors are statistical only.
The results for the $\sigma_1$, $\sigma_2$ masses and widths are given 
in the text.}

\label{dkpp_table}
\begin{tabular}{|l|c|c|c|c|c|} \hline
Intermediate state           & Amplitude & Phase ($^{\circ}$) \\ \hline

$K^*(892)^+\pi^-$            & $1.656\pm 0.012$
                             & $137.6\pm 0.6$ 
                             \\ 

$K^*(892)^-\pi^+$            & $(14.9\pm 0.7)\times 10^{-2}$
                             & $325.2\pm 2.2$
                             \\

$K_0^*(1430)^+\pi^-$         & $1.96\pm 0.04$
                             & $357.3\pm 1.5$
                             \\

$K_0^*(1430)^-\pi^+$         & $0.30\pm 0.05$
                             & $128\pm 8$
                             \\

$K_2^*(1430)^+\pi^-$         & $1.32\pm 0.03$
                             & $313.5\pm 1.8$
                             \\

$K_2^*(1430)^-\pi^+$         & $0.21\pm 0.03$
                             & $281\pm 9$
                             \\

$K^*(1680)^+\pi^-$           & $2.56\pm 0.22$
                             & $70\pm 6$
                             \\

$K^*(1680)^-\pi^+$           & $1.02\pm 0.2$
                             & $103\pm 11$
                             \\

$K_S\rho^0$                  & $1.0$ (fixed)                                 
                             & 0 (fixed)   
                             \\

$K_S\omega$                  & $(33.0\pm 1.3)\times 10^{-3}$
                             & $114.3\pm 2.3$
                             \\

$K_S f_0(980)$               & $0.405\pm 0.008$
                             & $212.9\pm 2.3$
                             \\

$K_S f_0(1370)$              & $0.82\pm 0.10$ 
                             & $308\pm 8$
                             \\

$K_S f_2(1270)$              & $1.35\pm 0.06$
                             & $352\pm 3$
                             \\

$K_S \sigma_1$               & $1.66\pm 0.11$
                             & $218\pm 4$
                             \\

$K_S \sigma_2$               & $0.31\pm 0.05$
                             & $236\pm 11$
                             \\

non-resonant                 & $6.1\pm 0.3$ 
                             & $146\pm 3$ 
                             \\ 
\hline
\end{tabular}
\end{table}

\begin{table}
\caption{Fractions of different background sources.}
\label{bck_table}
\begin{tabular}{|l|c|c|} \hline
Background source                     & \bdk & \bdsk \\ \hline
$q\bar{q}$ combinatorial              & $22.1\pm 3.9$\% & $9.0\pm 3.6$\%    \\
$B\bar{B}$ events other than $B^{\pm}\to D^{(*)}K^{\pm}/\pi^{\pm}$ 
                                      & $2.2\pm 0.2$\%  & $2.1\pm 0.4$\%    \\
\bddspi with $K/\pi$ misID            & $1.0\pm 0.2$\%  & $0.6\pm 0.2$\%    \\
Combinatorics in $D^0$ decay          & $0.4\pm 0.1$\%  & $0.4\pm 0.1$\%    \\
Combinatorial kaon in $B^{\pm}\to D^{(*)}K^{\pm}$ decay & $<$0.4\% (95\% CL) & $<$0.4\% (95\% CL) \\ \hline
Total                                 & $25\pm 4$\%     & $12\pm 4$\%     \\ \hline
\end{tabular}
\end{table}

\begin{table}
\begin{center}
\caption{Estimation of model uncertainty.}
\vspace{0.5\baselineskip}
\label{model_table}
\begin{tabular}{|c|c|c|c|c|} \hline
Fit model & $(\Delta r)_{\rm max}$ 
          & $(\Delta\phi_3)_{\rm max}$ ($^{\circ}$) 
          & $(\Delta\delta)_{\rm max}$ ($^{\circ}$) \\
\hline
$F_r=F_D=1$
    & 0.01 & 3.1 & 3.3 \\
$\Gamma(q^2)=Const$
    & 0.02 & 4.7 & 9.0 \\
Narrow resonances plus non-resonant term
    & 0.03 & 9.9 & 18.2 \\ \hline
Total
    & 0.04 & 11  & 21  \\
\hline
\end{tabular}
\end{center}
\end{table}

\begin{table}
\caption{Contributions to the experimental systematic error.}
\label{syst_table}
\begin{tabular}{|l|c|c|c|c|c|c|} \hline
                     & \multicolumn{3}{|c|}{\bdtk}
                     & \multicolumn{3}{|c|}{\bdstk} \\ 
		     \cline{2-7}
Source               & $\Delta r$ & $\Delta\phi_3$ ($^{\circ}$) & $\Delta\delta$ ($^{\circ}$) 
                     & $\Delta r$ & $\Delta\phi_3$ ($^{\circ}$) & $\Delta\delta$ ($^{\circ}$) \\ \hline
Background shape     & 0.017      & 4.7                  & 2.3       
                     & 0.016      & 1.5                  & 2.6                  \\
Background fraction  & 0.025      & 0.1                  & 0.6     
                     & 0.015      & 0.6                  & 0.9                  \\
Efficiency shape     & 0.004      & 3.5                  & 1.2    
                     & 0.002      & 3.5                  & 1.2                  \\
Momentum resolution  & 0.010      & 2.5                  & 0.6   
                     & 0.010      & 2.5                  & 0.6                  \\
Control sample bias  & 0.006      & 11                   & 11
                     & 0.006      & 11                   & 11                   \\ \hline 
Total                & 0.032      & 13                   & 11 
                     & 0.024      & 12                   & 11                   \\ \hline
\end{tabular}
\end{table}

\newpage

\end{document}